%% file: main.tex
\documentclass[review,12pt]{elsarticle}

\usepackage{latexsym,mathrsfs}
\usepackage{amsmath,amssymb}
\usepackage{amsthm,enumerate,verbatim}
\usepackage{amsfonts}
\usepackage{graphicx}
\usepackage[dvipsnames]{xcolor}
\usepackage{url}
\usepackage{pgfplots}
\pgfplotsset{compat=1.17}
\usepgfplotslibrary{statistics}
\usepgfplotslibrary{fillbetween}
\usepgfplotslibrary{groupplots}
\usepackage[hidelinks]{hyperref}
\usepackage{subcaption}
\usepackage{array}
\usepackage{diagbox}
\usepackage{cleveref}
\usepackage{enumitem}
\usepackage{booktabs}
\usepackage{multirow}
\usepackage{algpseudocode}
\usepackage[linesnumbered,ruled,vlined,commentsnumbered]{algorithm2e}
\newtheorem{lemma}{Lemma}

\newtheorem{theorem}{Theorem}
\newtheorem{corollary}{Corollary}

\newtheorem*{definition*}{Definition}

\newtheorem{assumption}{Assumption}

\DeclareMathOperator*{\argmin}{argmin}

\Crefname{figure}{Fig.}{Figs.}
\crefname{algorithm}{Alg.}{Algs.}
\Crefname{algorithm}{Algorithm}{Algorithms}

\newcommand{\revision}[1]{#1}

\newcommand{\ngi}[1]{#1}

\newcommand{\ngc}[1]{}

\newcommand{\revisionSP}[1]{{{\color{black} #1}}}
\newcommand{\revisionSPJ}[1]{{{\color{black} #1}}}

\journal{Signal Processing}

\begin{document}

\begin{frontmatter}

\title{Alternating Direction Method of Multipliers for\\
Nonlinear Matrix Decompositions}

\author[mons]{Atharva Awari\corref{cor1}}
\ead{atharva.awari05be@gmail.com}

\author[mons]{Nicolas Gillis}
\ead{nicolas.gillis@umons.ac.be}

\author[mons]{Arnaud Vandaele}
\ead{arnaud.vandaele@umons.ac.be}

\cortext[cor1]{Corresponding author.}

\affiliation[mons]{organization={Department of Mathematics and
    Operational Research, University of Mons},
  addressline={Rue de Houdain~9},
  postcode={7000},
  city={Mons},
  country={Belgium}}

\input{0-abstract}


\end{frontmatter}

\input{1-introduction}

\input{2-previous_works}

\input{3-motivations}

\input{5-admm_for_nmd}

\input{6-experiments}

\input{7-conclusion}

\section*{Acknowledgements}
The authors would like to thank L\^e Thi Khanh
Hi\^en for insightful discussions that initiated this project.
They are grateful to the anonymous reviewers who carefully read the manuscript and provided feedback that helped them improve the paper.

The authors acknowledge the support by the European Union (ERC
consolidator, eLinoR, no.\ 101085607), and by the F.R.S.-FNRS under the PDR project T.0097.22.

\section*{Declaration of competing interest}
The authors declare that they have no known competing financial interests or personal relationships that could have appeared to influence the work reported in this paper.

\section*{Declaration of generative AI and AI-assisted technologies in
the writing process} During the preparation of this work the authors used Claude (Anthropic)
to assist with clarifying and rephrasing sentences and with improving the readability of the manuscript. 
\revisionSP{We also used it to construct the proof of convergence of ADMM in the Supplementary.} 
All scientific content, core ideas, derivations and experimental results originate from the authors. After
using this tool, the authors reviewed and edited the content as needed and take full responsibility for the content of the publication.

\section*{Data availability}
The code implementing the proposed ADMM framework for all twelve models,
together with all scripts reproducing the experiments reported in this
paper, is publicly available at
\url{https://gitlab.com/Atharva05/admm-for-nmd}. All datasets used are
publicly available benchmarks.

\small 

\bibliographystyle{elsarticle-num}
\bibliography{ADMM2}

\normalsize
\input{supplementary}

\end{document}

%% file: 0-abstract.tex
\begin{abstract}
We present an algorithm based on the alternating direction method of multipliers (ADMM) for solving \emph{nonlinear matrix decompositions} (NMD). Given an input matrix $X \in \mathbb{R}^{m \times n}$ and a factorization rank $r \ll \min(m, n)$, NMD seeks matrices $W \in \mathbb{R}^{m \times r}$ and $H \in \mathbb{R}^{r \times n}$ such that $X \approx f(WH)$, where $f$ is an element-wise nonlinear function. We evaluate our method on several representative nonlinear models: the rectified linear unit activation $f(x) = \max(0, x)$, suitable for nonnegative sparse data approximation, the component-wise square $f(x) = x^2$, applicable to probabilistic circuit representation, and the MinMax transform $f(x) = \min(b, \max(a, x))$, relevant for recommender systems. The proposed framework flexibly supports diverse loss functions, including least squares, $\ell_1$ norm, and the Kullback-Leibler divergence, and can be readily extended to other nonlinearities and metrics. We illustrate the applicability, efficiency, and adaptability of the approach on real-world datasets, highlighting its potential for a broad range of applications.
\end{abstract}

%% file: 1-introduction.tex
\section{Introduction}

Low-rank matrix approximations (LRMAs) are key tools in data analysis and signal processing. When dealing with a data set stored in a matrix $X$, it is  customary to approximate it by the  product of two smaller matrices, called factors, 
and solve the following problem: Given $X \in \mathbb R ^ {m\times n}$ and $r\ll \min(m, n)$, find two factors, $W \in \mathbb{R}^{m \times r}$ and $H \in \mathbb{R}^{r \times n}$, such that $X \approx WH$. This is equivalent to linear dimensionality reduction, as each column of the matrix $X$ is approximated by a linear combination of the columns of $W$ , with the weights provided by the corresponding column of $H$:  for all $j$, 
\begin{equation*} \label{lindimred}
    X(:,j)
		\quad \approx \quad 
		\sum_{k=1}^r \; W(:,k) \; H(k,j).
\end{equation*}  
The primary objectives of LRMAs are twofold. First, by reducing the dimensionality of the data, LRMA enables more efficient computation in subsequent processing and analysis tasks. Second, it often facilitates the identification of the most informative features within the dataset, thereby supporting tasks such as pattern recognition and data interpretation. The most widely used and known LRMA is principal component analysis (PCA), which can be efficiently computed via the singular value decomposition (SVD) \cite{EckartYoung36}. 
Robust PCA \cite{candes2011robust}, sparse PCA \cite{dAspremont2007spca}, and PCA with weights or missing data \cite{srebro2003weighted} are some of the important variants of PCA.   %NMF, in contrast, imposes the additional constraint that both low-rank factors $W$ and $H$ are element-wise nonnegative. In this setting, when the data matrix $X$ is nonnegative, each entry of $W$ and $H$ remains nonnegative as well, yielding parts-based, additive representations that are particularly interpretable in applications such as image analysis, topic modeling, etc.

Although LRMAs are powerful and widely used, they are not universally applicable to all datasets or problem domains. As inherently linear models, LRMAs have limited expressivity. Many real-world datasets exhibit complex, nonlinear, and intertwined feature structures that cannot be adequately captured by linear models such as LRMAs. 
This limitation has motivated a growing interest in \emph{nonlinear matrix decompositions} (NMDs). First introduced by Saul~\cite{saul2022nonlinear}, NMDs generalize LRMAs to the nonlinear domain as follows 
\begin{equation*}
X \approx f(WH),
\end{equation*}
where $f$ is an element-wise nonlinear function. This framework enables capturing a richer and more intricate relationship within the data, making it well-suited for applications where the linearity assumption does not hold. 
\ngi{NMD is closely related to the post-nonlinear mixture model~\cite{taleb2002source} although, in this line of research, the non-linear function is not component-wise and is assumed to be unknown, which leads to rather different optimization problems; see~\cite{nguyendiverse} for a recent work on this topic.}  

\revisionSP{For example, NMDs have been used 
(i)~to compress sparse and dense
nonnegative data more efficiently than linear models~\cite{saul2022nonlinear,seraghiti2023accelerated,awari2024coordinate,lefebvre2024component}, see in particular the very recent paper~\cite{saul2026subzero},  
(ii)~for matrix completion with entries missing not at random~\cite{liu2024symmetric}, 
(iii)~to represent more expressive probabilistic circuits~\cite{loconte2025sum}, and 
(iv)~to model data bounded in an interval, such as ratings in recommender systems
or clipped signals~\cite{awari2026nonlinear}; these applications are described in Section~\ref{sec:background}.}

\revision{
\paragraph*{Contributions}

The main contributions of this paper are as follows:
We propose a unified and flexible ADMM framework for NMDs. We instantiate it for four nonlinear functions %(namely ReLU, square, MinMax and Modulus),   %including two new ones (namely, MinMax and Modulus), 
combined with three loss functions, for a total of 12 different models. %(Frobenius, $\ell_1$, KL divergence). %with closed-form  updates for all 12 combinations. What closed form? forwhat? Please do not used concept/ideas not explained yet. 
The framework 
is modular: incorporating a new function or loss requires only 
a scalar proximal operator, while the rest of the algorithm 
remains unchanged. We extend the framework to matrix with missing data. 
We provide comprehensive 
experiments showing that our proposed ADMM algorithm matches or outperforms 
specialized state-of-the-art methods on several tasks including matrix completion.   

\paragraph*{Outline}

The paper is organized as follows. 
Section~\ref{sec:background} provides background on NMDs 
and reviews existing nonlinear models. 
Section~\ref{sec:motivations} discusses the motivations 
behind the proposed framework. 
Section~\ref{sec:admm_nmd} describes the proposed ADMM 
algorithm for NMDs and derives the subproblem updates. 
Section~\ref{sec:experiments} presents numerical 
experiments on synthetic and real-world datasets.}

%% file: 2-previous_works.tex
\section{Background on NMDs}
\label{sec:background}
In recent years, several NMD models have attracted significant attention, among which the rectified linear unit–based NMD (ReLU-NMD) is particularly notable. The ReLU-NMD model was  introduced by Saul in~\cite{saul2022nonlinear}, 
it approximates a nonnegative and sparse matrix $X$ as
\begin{equation*}
X \approx \max(0, WH),
\end{equation*}
where the nonlinearity is the rectified linear unit $f(\cdot) = \max(0, \cdot)$, a function widely used as an activation in the hidden layers of neural networks. ReLU-NMD has been shown to be especially effective for compressing nonnegative sparse data. Notably, the identity matrix of any dimension can be exactly factorized by a rank-3 ReLU-NMD~\cite{saul2022nonlinear}. 
In a subsequent study, Saul further showed that ReLU-NMD can be interpreted as a ``faithful'' low-dimensional embedding of high-dimensional data~\cite{saul2022geometrical}. More recently,~\cite{seraghiti2023accelerated, gillis2025extrapolated} provided empirical evidence that ReLU-NMD achieves substantially better compression of sparse images compared to the SVD. ReLU-NMD can also be used for matrix completion. In particular, Liu et al.~\cite{liu2024symmetric} studied matrix completion with entries missing not at random (MNAR), where the probability of missing entries sometimes depends deterministically on the underlying values. They addressed this challenging setting using the ReLU-NMD framework to recover the missing entries. See also~\cite{gillis2025extrapolated} for an application on sensor network localization.

Another notable nonlinear model is the \emph{component-wise square factorization} (CSF), which decomposes a matrix $X$ as the Hadamard (element-wise) square of a low-rank matrix:
\begin{equation*}
    X \approx (WH)^{.2} = (WH) \circ(WH).
\end{equation*}
CSF provides a more expressive representation of probabilistic circuits; in particular, it was proven in~\cite{loconte2024subtractive} to be exponentially more expressive than monotonic probabilistic circuits. Furthermore, CSF can be used to compute the \emph{square-root rank}, a quantity relevant to the compact representation of convex polytopes~\cite{lee2014square,fawzi2015positive}. Empirical studies have also shown that CSF can outperform linear models such as the SVD  in compressing dense nonnegative data~\cite{lefebvre2024component}.

A third nonlinear matrix decomposition model is the \emph{MinMax} model, which is particularly well-suited for datasets whose values are bounded within an interval $[a,b]$:
\begin{equation*}
    X \approx \min(b, \max(a, WH)).
\end{equation*}
This formulation is closely related to existing approaches in nonlinear low-rank modeling. When $X$ is constrained to lie in $[a,b]$, the problem is connected to \emph{bounded matrix factorization} (BMF)~\cite{kannan2014bounded, thanh2023bounded}, a variant of traditional matrix factorization in which the predicted values are restricted to a fixed range. Such constraints are particularly relevant in applications like recommender systems, where ratings or preferences naturally fall within a predefined scale (e.g., 1--5 stars). Typical examples include the \emph{MovieLens}~\cite{harper2015movielens} and the \emph{Netflix Prize} datasets~\cite{bennett2007netflix}. 

\revisionSP{The MinMax model has recently been used in tensor settings for clipped data, where values beyond a known threshold are missing, for instance, in saturated  audio recordings~\cite{kitic2014audio} or in metabolomics~\cite{wei2018missing}, where values below the detection limit of the instrument are not measured. Unlike linear models, which either ignore the clipped entries or treat the threshold as their true value, the MinMax model uses the threshold as a bound on the missing entries and only requires them to lie beyond the threshold. This idea has recently been used  with a nonlinear CP decomposition based on the MinMax function, for pattern discovery in clipped metabolomics data~\cite{awari2026nonlinear}.}

A fourth nonlinear matrix decomposition model that we introduce in this paper is the \emph{modulus} model, defined as $X \approx |WH|$. It would be interesting to compare it to CSF to approximate nonnegative data, such as dense images.

The models described above are not intended to be an exhaustive list; other nonlinear formulations exist in the literature, either directly or indirectly related to the approaches considered here, and additional variants are likely to emerge in the future. In this paper, however, we restrict our focus to the nonlinear models described above.

%% file: 3-motivations.tex
\section{Motivations}
\label{sec:motivations}
The motivations for proposing an ADMM algorithm for NMDs 
are fourfold: 

\noindent a) The literature contains a variety of nonlinear models, and with the growing interest in NMDs, many more are likely to emerge. Dedicated algorithms have been developed to address specific instances, such as ReLU-NMD \cite{saul2022nonlinear, seraghiti2023accelerated, awari2024coordinate} and CSF-NMD \cite{lefebvre2024component}. However, to the best of our knowledge, no algorithms currently exist for the MinMax-NMD or the Modulus-NMD. Furthermore, there is no general algorithmic framework capable of simultaneously handling multiple nonlinear models in a unified and flexible manner. In this work, we aim to address these gaps by introducing an algorithm designed to accommodate a broad class of nonlinear models within a single framework. As mentioned before, we will consider in this paper the following nonlinear functions: 
\begin{itemize}
    \item \textbf{ReLU:} $f(T) = \max(0, T)$, 
    \item \textbf{CSF:} $f(T) = T^{.2}$,
    \item \textbf{MinMax:} $f(T) = \min(b, \max(a, T))$,
    \item \textbf{Modulus:} $f(T) = |T|$.   
\end{itemize}  

\noindent b) In matrix decompositions, the goal is to approximate a data matrix $X \in \mathbb R^{m \times n}$ with a low-rank matrix $\Tilde{X}\in \mathbb R^{m \times n}$. The optimization problem is  formulated as 
    \begin{equation}
        \min_{\Tilde{X}} d(X,\Tilde{X}),
        \label{matrix_decom}
    \end{equation}
    where $d: \mathbb R^{m \times n} \times \mathbb R^{m \times n} \rightarrow \mathbb R_{\geq 0}$  is the \emph{loss function} which measures the discrepancy between the original matrix $X$ and its low-rank approximation $\Tilde{X}$. 
    \revision{We will assume that the objective function $d(\cdot,\cdot)$ is separable, that is, $d(X,\Tilde{X}) = \sum_{i,j} d_{i,j}(X_{i,j},\Tilde{X}_{i,j})$ for some scalar functions $d_{i,j}(\cdot,\cdot)$. In this paper, the scalar functions, $d_{i,j}(\cdot,\cdot)$, will be the same for all $(i,j)$, but this is not necessary to apply our framework.}  
    
    The most commonly used loss function is the Frobenius norm (least-squares loss), which corresponds to the maximum likelihood estimator under the assumption of additive, independent Gaussian noise in the entries of $X$. Its differentiability facilitates efficient algorithm design for solving \eqref{matrix_decom}. However, the Frobenius norm is not always the most appropriate choice. For data with non-Gaussian noise, alternative loss functions can yield more meaningful factorizations. For instance, the $\ell_1$ loss is robust to outliers, while the Kullback--Leibler (KL) divergence is better suited for data following the Poisson distribution.  

    Currently, no algorithms for NMDs exist that can accommodate  loss functions other than the Frobenius norm; in particular   algorithms for ReLU-NMD and CSF handle only the Frobenius norm. To address this gap, our ADMM-based algorithm is capable of handling NMDs with other loss functions; in this paper we consider the following ones:  
\begin{itemize}
    \item \textbf{Frobenius norm:} $\|X\text{$-$}f(T)\|^2_F$~where~\mbox{$\|A\|_F^2 = \sum_{i,j} A_{ij}^2$}.
    \item \textbf{$\ell_1$ norm:} \mbox{$\|X - f(T)\|_1 = \sum_{i,j} \left|\, X_{ij} - f(T)_{ij} \,\right|$}. 
    
    \item \textbf{Kullback--Leibler (KL) divergence:} 
    \[
        KL(X , f(T)) = \sum_{i,j} d_{\text{KL}} (X_{ij}, f(T)_{ij}) , 
    \]  
    where, for two nonnegative scalars $x$ and $y$,  
    \[
        d_{\text{KL}}(x,y) = 
        \begin{cases}
            x \log \!\left( \tfrac{x}{y} \right) - x + y & \text{if } x > 0, \\
            y & \text{if } x = 0. 
        \end{cases}
    \]
\end{itemize}

\noindent c) With our proposed algorithm, our aim is to provide flexible support for a wide range of combinations between NMD models and loss functions. Specifically, users can pair different NMD variants, such as ReLU-NMD or MinMax-NMD, with various loss functions. %including the Frobenius norm, $\ell_1$ norm, or Kullback--Leibler (KL) divergence. 
For example, it becomes possible to perform ReLU-NMD with the KL divergence or MinMax-NMD with the $\ell_1$ norm, and many other combinations can be explored depending on the characteristics of the data and the underlying noise present. 
In particular, our ADMM framework will be able to handle the 4 nonlinear functions with the 3 loss functions described above, for a total of 12 models. 
To the best of our knowledge, this is the first algorithmic framework that introduces such versatility, allowing a unified and flexible approach to NMDs. This flexibility significantly broadens the applicability of NMDs across diverse datasets and problem domains, enabling more accurate modeling under varying noise conditions and structural constraints.

\noindent d) As discussed previously, the growing interest in NMDs, combined with empirical evidence showing that NMDs are more expressive than traditional LRMAs, suggests that a wide variety of new nonlinear models are likely to be explored in the future. A key goal of our work is to ensure that the proposed algorithm remains future proof, capable of accommodating these emerging models. Thanks to the design of our ADMM-based framework, additional nonlinear models can be incorporated easily into the existing structure. 
Our approach is not only highly flexible, capable of handling a wide range of currently studied NMDs, but also adaptable, providing a foundation for integrating novel nonlinear models as they are developed in future research.

%% file: 5-admm_for_nmd.tex
\section{ADMM for NMD}
\label{sec:admm_nmd}

As discussed earlier, we draw inspiration from ADMM for convex optimization, see \cite{boyd2011distributed} for a comprehensive treatment, and adapt it to the setting of NMDs. 
ADMM-based approaches \revision{are} a standard choice for low-rank matrix approximations~\cite{sun2014alternating, li2016alternating}. For example, the \revision{AO-ADMM} framework proposed by Huang et al. \cite{huang2016flexible} for constrained matrix and tensor factorizations updates one factor at a time, solving the subproblems for each factor using ADMM. Their method supports a broad range of constraints and loss functions. \revision{In contrast, we apply ADMM directly to the full 
NMD problem over all variables $(W, H, T)$ 
simultaneously, rather than using it as an inner 
solver within an alternating optimization scheme}. The ADMM framework provides a natural way to handle NMDs across a variety of nonlinear models combined with different loss functions, as highlighted in the previous section. 

We now formally define the NMD optimization problem. Given a data matrix $X \in \mathbb{R}^{m \times n}$ and a factorization rank $r \ll \min(m,n)$, the goal is to solve
\begin{equation}
\min_{W,H,T} \; d\big(X, f(T)\big) \quad \text{subject to} \quad T = WH,
\label{eq:nmd_formulation}
\end{equation}
where $T \in \mathbb{R}^{m \times n}$ is a rank-$r$ matrix representing a low-rank approximation of $X$, $W \in \mathbb{R}^{m \times r}$ and $H \in \mathbb{R}^{r \times n}$ are the rank-$r$ factors, $f(\cdot)$ is an element-wise nonlinear function, and $d(\cdot, \cdot)$ is the chosen loss function. 
By parameterizing the low-rank approximation as $T = WH$, we avoid the explicit rank constraint $\text{rank}(T) = r$, which is generally more difficult to enforce directly. This factorized representation aligns naturally with the ADMM framework.

The augmented Lagrangian for NMD~\eqref{eq:nmd_formulation} is  
\begin{equation}
    L(W,H,T,\Lambda) = d(X, f(T)) + \langle  T - WH \,, \Lambda \rangle + \frac{\rho}{2} \|T - WH\|_F^2, \label{eq:aug_lag}
\end{equation}  
where $\rho > 0$ is a {penalty parameter}. 
The ADMM procedure alternates between minimizing the augmented Lagrangian with respect to the primal variables $W, H$, and $T$, followed by a dual variable update for $\Lambda$. 
The iterations are given by  
\begin{subequations} \label{eq:nmd_admm_updates}
\begin{align}
    W^{k+1} &:= \argmin_W \; L(W, H^k, T^k, \Lambda^k), \label{eq:nmd_W} \\
    H^{k+1} &:= \argmin_H \; L(W^{k+1}, H, T^k, \Lambda^k), \label{eq:nmd_H} \\
    T^{k+1} &:= \argmin_T \; L(W^{k+1}, H^{k+1}, T, \Lambda^k), \label{eq:nmd_T} \\
    \Lambda^{k+1} &:= \Lambda^{k} + \rho \,\big(T^{k+1} - W^{k+1}H^{k+1}\big). \label{eq:nmd_Lambda}
\end{align}
\end{subequations}  
In the following, we describe how to update each block of variables.

\subsection{Updating factors $W$ and $H$} 

The updates for the factors $W$ and $H$
correspond to the minimization steps~\eqref{eq:nmd_W} and~\eqref{eq:nmd_H}, resp.
Let us first consider the update for~$W$. 
Since the loss term $d(X,f(T))$ does not depend on~$W$,
the subproblem in $W$ simplifies to
\begin{equation*}
\label{eq:admm_W_simple}
W^{k+1}
= \argmin_W
\Big\{
\langle T - WH,\; \Lambda \rangle 
+ \tfrac{\rho}{2}\,\|T - WH\|_F^2
\Big\}, 
\end{equation*}
where we removed the iteration index $k$ for simplicity of the presentation. 
We simplify the objective using 
$\|A+B\|_F^2 = \|A\|_F^2 + 2\langle A,B\rangle + \|B\|_F^2$:
\begin{align*}
\label{eq:admm_W_square}
\langle T - WH,\, \Lambda \rangle & + \tfrac{\rho}{2}\|T - WH\|_F^2 \\  
&= \tfrac{\rho}{2}   \Big\|(T - WH) + \tfrac{\Lambda}{\rho}\Big\|_F^2 - \tfrac{1}{2\rho}\,\|\Lambda\|_F^2 .
\end{align*}
The last term is constant with respect to~$W$, hence 
\begin{equation}
\label{eq:admm_W_ls}
\begin{aligned}
W^{k+1}
&= \argmin_W
   \Big\|(T^{k} - W H^{k}) + \tfrac{\Lambda^{k}}{\rho}\Big\|_F^2. 
\end{aligned}
\end{equation}
The first-order optimality condition of~\eqref{eq:admm_W_ls} yields
\begin{equation*}
W^{k+1}
= \Big(T^{k} + \tfrac{\Lambda^{k}}{\rho}\Big)
  H^{k\top}(H^{k}H^{k\top})^{-1}.
\end{equation*}
In practice, however, $H^{k}H^{k\top}$ can be ill-conditioned or singular, so we instead use a ridge-regularized formulation \cite{tikhonov1963regularization}, which is both more stable and always well-defined:
\begin{equation}
\label{eq:admm_W_ridge}
W^{k+1} \; = \; \Big(T^{k} + \tfrac{\Lambda^{k}}{\rho}\Big) \,H^{k\top} \, 
   \big(H^{k}H^{k\top} + \varepsilon_W I_r\big)^{-1}, 
   \end{equation} 
   where  $\varepsilon_W = 10^{-6}\,\|H^{k}\|_F^2$ and 
$I_r$ is the identity matrix of dimension $r$.

By symmetry, the update for~$H$ can be derived analogously, leading to the closed-form expression: 
\begin{equation} \label{eq:admm_H_closed} 
H^{k+1} \; = \;  \big(W^{(k+1)\top} W^{(k+1)} + \varepsilon_H I_r\big)^{-1}
   W^{(k+1)\top} \Big(T^k + \frac{\Lambda^k}{\rho}\Big), 
   \end{equation} 
where $\varepsilon_H = 10^{-6}\,\|W^{(k+1)}\|_F^2$. 
Both factor updates, therefore, correspond to
ridge-regularized quadratic minimization problems,
independent of the specific nonlinear function~$f(\cdot)$
or loss function~$d(\cdot,\cdot)$.
This underlines a key advantage of our framework:
the updates for $W$ and $H$ retain the same closed-form structure
regardless of the choice of nonlinearity or loss,
making the algorithm broadly flexible and easily adaptable
to different modeling choices.

\subsection{Updating the matrix $T$}  \label{app:relu_frob} 

The update for $T$ is obtained by minimizing the augmented Lagrangian with respect to $T$: 
\[
T^{k+1} \in \argmin_T \, d(X, f(T)) + \langle T - WH, \Lambda \rangle + \tfrac{\rho}{2}\|T - WH\|_F^2 .
\] 
Since the objective is separable across the entries of $T$, each element can be updated independently. 
Let $x$, $t$, $\lambda$, and $a$ denote the $(i,j)$th entries of $X$, $T$, $\Lambda$, and  $A := WH$, respectively. 
The corresponding scalar subproblem is 
\[
T_{ij}^{k+1} := \argmin_t \, d(x, f(t)) + t \lambda + \tfrac{\rho}{2}(t - a)^2 .
\] 
Because the loss term $d(x,f(t))$ is nonnegative and the quadratic penalty term is strictly convex when $\rho > 0$, the objective is coercive (it goes to infinity as $t$ goes to $\pm \infty$), hence the global minimum is attained (although it might not be unique). %
In cases where the chosen loss function $d$ and nonlinearity $f$ are  simple enough, the global minimum can be expressed in a closed analytical form. This will be the case for the 12 models we consider in this paper. \revision{For more complicated nonlinearities or loss functions, 
the update may no longer admit a closed analytical form. 
However, since the $T$-update decouples into independent 
univariate subproblems and the quadratic penalty ensures 
coercivity, efficient root-finding methods (e.g.,  
bisection, Newton's method) can be used to tackle the subproblems.}

Importantly, this step is the only one in the algorithm that depends on the choice of the nonlinearity, $f(\cdot)$, and the loss function, $d(\cdot,\cdot)$. 
In contrast, the updates for $W$, $H$, and $\Lambda$ are  model independent. 
This distinction highlights a central strength of our framework: the algorithm remains flexible and broadly applicable, with all model-specific complexity isolated to the $T$-update.

\begin{algorithm}[htb!]
\caption{ADMM for NMD: General Framework}
\label{alg:admm_nmd}
\SetKwInOut{Input}{Input}
\SetKwInOut{Output}{Output}
\DontPrintSemicolon

\Input{ 
    Data matrix $X \in \mathbb{R}^{m \times n}$, target rank $r$, penalty parameter $\rho$, maximum iterations $K$, maximum runtime. 
} 

\Output{ 
    Factor matrices $W \in \mathbb{R}^{m \times r}$ and $H \in \mathbb{R}^{r \times n}$ such that $X \approx f(WH)$.
}

\BlankLine
\textbf{Initialization:} 
$W \in \mathbb{R}^{m \times r}, \;
 H \in \mathbb{R}^{r \times n}, \;
 T \in \mathbb{R}^{m \times n}, \;
 \Lambda \in \mathbb{R}^{m \times n}$.

\BlankLine
\While{iteration $< K$ \textbf{ and } runtime $<$ max runtime}{
    
    \BlankLine
    \textbf{Update $W$:} for  $\varepsilon_W = 10^{-6}\,\|H^{k}\|_F^2$, let 
    \[
    W^{k+1} \;\; \gets \;\; 
    \tfrac{1}{\rho}\, 
    \big(\Lambda H^\top + \rho \, T H^\top \big)(H H^\top +\varepsilon_W I_r)^{-1}. 
    \]
    
    \BlankLine
    \textbf{Update $H$:} for  $\varepsilon_H = 10^{-6}\,\|W^{k+1}\|_F^2$, let 
    \[
    H^{k+1} \;\; \gets \;\;
    \tfrac{1}{\rho}\,(W^\top W + \varepsilon_H I_r)^{-1} 
    \big(W^\top \Lambda + \rho \, W^\top T \big). 
    \] 
    
    \BlankLine
    \textbf{Update $T$:} Solve $mn$ entrywise subproblems:
    \[
    t_{ij}^{k+1} \;\; \in \;\; 
    \argmin_{t} \;
    d(X_{ij}, f(t)) + t \Lambda_{ij}^k + \tfrac{\rho}{2}(t - A_{ij})^2 , 
    \]
    where $A= W^{k+1} H^{k+1}$. 
    
    \BlankLine
    \textbf{Update dual variable $\Lambda$:} 
    \[
    \Lambda^{k+1} \;\; \gets \;\; 
    \Lambda^k + \rho \big(T^{k+1} - W^{k+1}H^{k+1}\big). 
    \]
}
\end{algorithm}

\revisionSP{\paragraph{Computational cost} 
Each iteration of Algorithm~\ref{alg:admm_nmd} requires $\mathcal{O}(mnr)$ operations.  The cost is dominated by the matrix products $(\Lambda + \rho T)H^\top$, $W^\top(\Lambda + \rho T)$ and $WH$.  The $r \times r$ linear systems in the $W$ and $H$ updates cost $\mathcal{O}(r^3 + (m+n) r^2)$ operations, which is negligible since $r \ll \min(m,n)$.  
The $T$ update requires the computation of $A = W^{k+1}H^{r+1}$ in $\mathcal{O}(mnr)$ operations, and then requires solving $mn$ one-dimensional subproblems which, in all 12 cases, require $\mathcal{O}(1)$ operations. 
The $\Lambda$ updates requires $\mathcal{O}(mn)$ additional operations since $A$ is already computed. 
Algorithm~\ref{alg:admm_nmd} stores 4 $m \times n$ matrices ($X$, $T$, $\Lambda$ and $WH$), requiring $\mathcal{O}(mn)$ memory, which is of the same order as the data.  The same cost applies in the presence of missing data, since only the $T$-update changes. 
}

To illustrate the update of the matrix $T$, we present the case of the ReLU nonlinearity with the Frobenius norm in the next section. 
All other combinations are presented in Supplementary~S1.

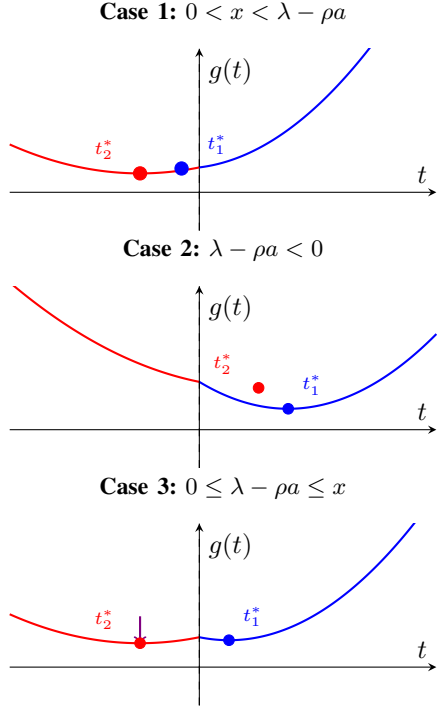
\begin{figure}[htbp]
\centering
\begin{minipage}[t]{0.32\textwidth}
\centering
\begin{tikzpicture}
  \begin{axis}[
    width=0.95\linewidth, height=0.85\linewidth,
    axis x line=middle, axis y line=middle,
    xmin=-1.6, xmax=2.0, ymin=-0.8, ymax=3.0,
    xtick={0}, ytick=\empty,
    xlabel={$t$}, ylabel={$g(t)$},
    xlabel style={font=\scriptsize, at={(ticklabel* cs:1.0)}, anchor=west},
    ylabel style={font=\scriptsize, at={(ticklabel* cs:1.0)}, anchor=south},
    tick label style={font=\tiny},
    title={\textbf{Case 1}}, title style={font=\small, yshift=4pt},
  ]
    \def\xval{0.2}\def\aval{1.0}\def\lam{1.5}\def\rhoa{1.0}
    \addplot[domain=-1.6:0, samples=120, thick, red]
      {0.5*(\xval)^2 + \lam*x + 0.5*\rhoa*(x-\aval)^2};
    \addplot[domain=0:2.0, samples=120, thick, blue]
      {0.5*(\xval - x)^2 + \lam*x + 0.5*\rhoa*(x-\aval)^2};
    \pgfmathsetmacro{\tone}{(\xval + \rhoa*\aval - \lam)/(\rhoa+1)}
    \pgfmathsetmacro{\ttwo}{(\rhoa*\aval - \lam)/\rhoa}
    \pgfmathsetmacro{\gone}{0.5*(\xval-\tone)^2 + \lam*\tone + 0.5*\rhoa*(\tone-\aval)^2}
    \pgfmathsetmacro{\gtwo}{0.5*(\xval)^2 + \lam*\ttwo + 0.5*\rhoa*(\ttwo-\aval)^2}
    \addplot[only marks, mark=*, mark size=1.6pt, blue] coordinates {(\tone,\gone)};
    \addplot[only marks, mark=*, mark size=1.6pt, red]  coordinates {(\ttwo,\gtwo)};
    \node[blue, font=\tiny, anchor=south west] at (axis cs:\tone,\gone) {$t_1^*$};
    \node[red,  font=\tiny, anchor=south east] at (axis cs:\ttwo,\gtwo) {$t_2^*$};
  \end{axis}
\end{tikzpicture}
\end{minipage}
\hfill
\begin{minipage}[t]{0.32\textwidth}
\centering
\begin{tikzpicture}
  \begin{axis}[
    width=0.95\linewidth, height=0.85\linewidth,
    axis x line=middle, axis y line=middle,
    xmin=-1.6, xmax=2.0, ymin=-0.8, ymax=3.0,
    xtick={0}, ytick=\empty,
    xlabel={$t$}, ylabel={$g(t)$},
    xlabel style={font=\scriptsize, at={(ticklabel* cs:1.0)}, anchor=west},
    ylabel style={font=\scriptsize, at={(ticklabel* cs:1.0)}, anchor=south},
    tick label style={font=\tiny},
    title={\textbf{Case 2}}, title style={font=\small, yshift=4pt},
  ]
    \def\xval{1.0}\def\aval{1.0}\def\lam{0.5}\def\rhoa{1.0}
    \addplot[domain=-1.6:0, samples=120, thick, red]
      {0.5*(\xval)^2 + \lam*x + 0.5*\rhoa*(x-\aval)^2};
    \addplot[domain=0:2.0, samples=120, thick, blue]
      {0.5*(\xval - x)^2 + \lam*x + 0.5*\rhoa*(x-\aval)^2};
    \pgfmathsetmacro{\tone}{(\xval + \rhoa*\aval - \lam)/(\rhoa+1)}
    \pgfmathsetmacro{\ttwo}{(\rhoa*\aval - \lam)/\rhoa}
    \pgfmathsetmacro{\gone}{0.5*(\xval-\tone)^2 + \lam*\tone + 0.5*\rhoa*(\tone-\aval)^2}
    \pgfmathsetmacro{\gtwo}{0.5*(\xval)^2 + \lam*\ttwo + 0.5*\rhoa*(\ttwo-\aval)^2}
    \addplot[only marks, mark=*, mark size=1.6pt, blue] coordinates {(\tone,\gone)};
    \addplot[only marks, mark=*, mark size=1.6pt, red]  coordinates {(\ttwo,\gtwo)};
    \node[blue, font=\tiny, anchor=south west] at (axis cs:\tone,\gone) {$t_1^*$};
    \node[red,  font=\tiny, anchor=south east] at (axis cs:\ttwo,\gtwo) {$t_2^*$};
  \end{axis}
\end{tikzpicture}
\end{minipage}
\hfill
\begin{minipage}[t]{0.32\textwidth}
\centering
\begin{tikzpicture}
  \begin{axis}[
    width=0.95\linewidth, height=0.85\linewidth,
    axis x line=middle, axis y line=middle,
    xmin=-1.6, xmax=2.0, ymin=-0.8, ymax=3.0,
    xtick={0}, ytick=\empty,
    xlabel={$t$}, ylabel={$g(t)$},
    xlabel style={font=\scriptsize, at={(ticklabel* cs:1.0)}, anchor=west},
    ylabel style={font=\scriptsize, at={(ticklabel* cs:1.0)}, anchor=south},
    tick label style={font=\tiny},
    title={\textbf{Case 3}}, title style={font=\small, yshift=4pt},
  ]
    \def\xval{1.0}\def\aval{0.5}\def\lam{1.0}\def\rhoa{1.0}
    \addplot[domain=-1.6:0, samples=120, thick, red]
      {0.5*(\xval)^2 + \lam*x + 0.5*\rhoa*(x-\aval)^2};
    \addplot[domain=0:2.0, samples=120, thick, blue]
      {0.5*(\xval - x)^2 + \lam*x + 0.5*\rhoa*(x-\aval)^2};
    \pgfmathsetmacro{\tone}{(\xval + \rhoa*\aval - \lam)/(\rhoa+1)}
    \pgfmathsetmacro{\ttwo}{(\rhoa*\aval - \lam)/\rhoa}
    \pgfmathsetmacro{\gone}{0.5*(\xval-\tone)^2 + \lam*\tone + 0.5*\rhoa*(\tone-\aval)^2}
    \pgfmathsetmacro{\gtwo}{0.5*(\xval)^2 + \lam*\ttwo + 0.5*\rhoa*(\ttwo-\aval)^2}
    \addplot[only marks, mark=*, mark size=1.6pt, blue] coordinates {(\tone,\gone)};
    \addplot[only marks, mark=o, mark size=3pt, very thick, red]
      coordinates {(\ttwo,\gtwo)};
    \addplot[only marks, mark=*, mark size=1.6pt, red]  coordinates {(\ttwo,\gtwo)};
    \node[blue, font=\tiny, anchor=south west] at (axis cs:\tone,\gone) {$t_1^*$};
    \node[red,  font=\tiny, anchor=south east] at (axis cs:\ttwo,\gtwo) {$t_2^*$};
  \end{axis}
\end{tikzpicture}
\end{minipage}

\caption{Piecewise quadratics for the $T$-update. Blue: $g_1(t)$ for $t>0$, Red: $g_2(t)$ for $t\le 0$. All cases show $t_1^*$ and $t_2^*$.}
\label{fig:T-update-cases-final}
\end{figure}

\paragraph*{Update of $T$ with ReLU and Frobenius norm}  
The $T$-update at iteration $k$ is obtained by solving
\begin{equation}
\label{eq:T-update}
\argmin_{T \in \mathbb{R}^{m \times n}} \;  
\frac{1}{2}\,\|X - \max(0,T)\|_F^2  
+ \langle T, \Lambda^k \rangle
+ \frac{\rho}{2}\,\|T - A^{k+1}\|_F^2, 
\end{equation}
where $A^{k+1} = W^{k+1}H^{k+1}$. %and the nonlinearity $f(\cdot) = \max(0,\cdot)$ is applied entrywise.
Since the objective in \eqref{eq:T-update} is separable across entries of $T$, the problem reduces to solving $mn$ one-dimensional subproblems. 
Let us fix $(i,j)$, let $x = X_{ij} \geq 0$, $a = A^{k+1}_{ij}$, and $\lambda = \Lambda^k_{ij}$. 
The scalar problem is
\begin{equation*}
t^{k+1} \;\in\; \argmin_{t \in \mathbb{R}}
\;\frac{1}{2}\,\big(x - \max(0,t)\big)^2
\;+\; \lambda t
\;+\; \frac{\rho}{2}(t-a)^2.
\end{equation*}
This splits into two quadratic functions:
\[
g(t) = 
\left\{ \hspace{-0.4cm} 
\begin{array}{cll}
  & g_1(t)   := \tfrac{1}{2}(x-t)^2 + \lambda t + \tfrac{\rho}{2}(t-a)^2,  & \text{ for } t>0,     \\
   & g_2(t)  := \tfrac{1}{2}x^2 + \lambda t + \tfrac{\rho}{2}(t-a)^2, \quad & \text{ for } t\le 0. 
\end{array} 
\right. 
\] 
Their unconstrained minimizers are
\[
t_1^* \;=\; \frac{x + \rho a - \lambda}{\rho+1},
\qquad
t_2^* \;=\; \frac{\rho a - \lambda}{\rho}.
\] 
Using the thresholds implied by $t_1^*>0 \Leftrightarrow \lambda - \rho a < x$ and $t_2^*\le 0 \Leftrightarrow \lambda - \rho a \ge 0$, the scalar update $t^{k+1}$ can be described in three distinct cases corresponding to the relative positions of the two minimizers, $t_1^*$ and $t_2^*$, with respect to zero: 
\begin{enumerate}
    \item \textbf{Case 1:} $t_2^* < t_1^* < 0$ (i.e., $\lambda - \rho a > x$): $t^{k+1} = t_2^*$.

    \item \textbf{Case 2:} $0 < t_2^* < t_1^*$ (i.e., $\lambda - \rho a < 0$): $t^{k+1} = t_1^*$.

    \item \textbf{Case 3:} $t_2^* \leq 0 \leq t_1^*$ (i.e., $0 \leq \lambda - \rho a \leq x$):  
    $t^{k+1} = t_1^*$ if $g_1(t_1^*) < g_2(t_2^*)$, and 
    $t^{k+1} = t_2^*$ otherwise. 
\end{enumerate} 
These cases are illustrated on Figure~\ref{fig:T-update-cases-final}, highlighting the position of the candidate minimizers and the selected $t^{k+1}$.

Finally, the ReLU–Frobenius $T$-update has a closed form, selecting between two quadratic minimizers according to simple sign/threshold tests. 
Algorithm \ref{alg:admm_v1} provides a detailed description of our proposed ADMM algorithm for the ReLU nonlinearity combined with the Frobenius norm.

\begin{algorithm}[htbp]
\setcounter{AlgoLine}{0}
\caption{Update of $T$ for ReLU + Frobenius}
\label{alg:admm_v1}
\SetKwInOut{Input}{Input}
\SetKwInOut{Output}{Output}
\DontPrintSemicolon

\Input{ 
    Data matrix $X \in \mathbb{R}_+^{m \times n}$, 
    penalty parameter $\rho$, 
    $A = WH \in \mathbb{R}^{m \times n}, \;
 \Lambda \in \mathbb{R}^{m \times n}$.
} 

\Output{ 
    Matrix $T$ solving 
    \eqref{eq:T-update}.  
}
   Let $Y \;\gets\; \Lambda - \rho A$.

 Define the minimizers:  \[
 T^1 = \frac{1}{1+\rho}\bigl( X + \rho A - \Lambda \bigr),
 \qquad
 T^2 = A - \frac{1}{\rho}\Lambda .
 \]
   
Partition index sets:
     \[
     \begin{aligned}
     \Gamma &\gets \{\, (i,j) : Y_{ij} \leq 0 \,\}, \\
     \Delta &\gets \{\, (i,j) : Y_{ij} \geq X_{ij} \,\}, \\
     \Theta &\gets \{\, (i,j) : 0 < Y_{ij} < X_{ij} \,\}.
     \end{aligned}
     \]
   \vspace{0.2cm}
   
   Update $T$ entrywise: $T(\Gamma) = T^1(\Gamma)$, $T(\Delta) = T^2(\Delta)$. \vspace{0.2cm}  
    
    For the ambiguous region $\Theta$, we evaluate, for each $(i,j)\in\Theta$, the objective values: 
\[
\begin{aligned}
G_{{ij}}^1 
&=
\frac12 \bigl(  X_{ij} - T_{{ij}}^1 \bigr)^2
+ \Lambda_{ij} T_{{ij}}^1  + 
\frac{\rho}{2}
\bigl( T_{{ij}}^1 - A_{ij}\bigr)^2, \\ 
G_{{ij}}^2
&=
\frac12 \bigl(  X_{ij} \bigr)^2
+ \Lambda_{ij} T_{{ij}}^2  + 
\frac{\rho}{2}
\bigl( T_{{ij}}^2 - A_{ij}\bigr)^2,
\end{aligned}
\]

Select the minimizer:
\[
T_{ij}=
\begin{cases}
T_{{ij}}^1, & \text{if } G_{{ij}}^1\le G_{{ij}}^2, \\[4pt]
T_{{ij}}^2, & \text{otherwise}.
\end{cases}
\quad \text{for all } (i,j)\in\Theta.
\]
\end{algorithm}

For the other 11 choices of nonlinearities and loss functions, we present in Supplementary S1, the corresponding closed-form updates, without delving into the full algorithmic details. Table~\ref{tab:Tupdates} provides the location of each update for the 12 cases. 

\begin{table}[htb]
\centering
\caption{Where to find the update rules for $T$ under different
nonlinearities and loss functions. Entries prefixed with~S refer to
sections of the supplementary material.}
\label{tab:Tupdates}
\footnotesize
\setlength{\tabcolsep}{5pt}
\begin{tabular}{@{}lccc@{}}
\toprule
\textbf{Nonlinearity} & \textbf{Frobenius} & \textbf{KL} & $\boldsymbol{\ell_1}$ \\
\midrule
\textbf{ReLU:}, $f(t)=\max(0,t)$            & Section~\ref{app:relu_frob} & S1.4 & S1.8 \\
\textbf{CSF:}, $f(t)=t^{2}$                 & S1.1 & S1.5 & S1.9 \\
\textbf{MinMax:}, $f(t)=\min(b,\max(a,t))$  & S1.2 & S1.6 & S1.10 \\
\textbf{Modulus:}, $f(t)=|t|$               & S1.3 & S1.7 & S1.11 \\
\bottomrule
\end{tabular}
\end{table}

\subsection{Adaptive Update of the Penalty Parameter}
\label{subsection:adaptive}
The performance of ADMM is highly sensitive to the choice of the penalty parameter $\rho$. 
To reduce this sensitivity, it is beneficial to allow $\rho$ to vary across iterations, 
using an adaptive update scheme for $\rho^k$.

Following the idea from \cite{boyd2011distributed}, we propose an adaptive strategy relying on primal and dual feasibility. 

The dual function associated with the NMD problem 
\begin{equation*}
    \min_{W,H,T} \; d(X, f(T))
    \quad \text{s.t.} \quad T = WH,
\end{equation*}
is given by  
\begin{equation*}
    q(\Lambda) = \min_{W,H,T} 
    \; d(X, f(T)) + \langle T - WH, \Lambda \rangle.
\end{equation*}
The optimality conditions are: 
\begin{itemize}
    \item Primal feasibility: $T - W H = 0$.

    \item Dual feasibility: 
\begin{align}
    0 &\in \partial \big[d(X, f(T^*))\big] + \Lambda^*, 
    \label{eq:dual_feasibility} \\
    0 &= \Lambda^* H^{*\top}, 0 = W^{*\top} \Lambda^*, 
    \label{eq:dual_feasibility2}
\end{align}
where $\partial$ denotes the subdifferential. If $d$ and $f$ are differentiable, it reduces to the gradient. 
\end{itemize}

The augmented Lagrangian is
\begin{equation*}
    L_\rho(W,H,T,\Lambda) 
    = d(X, f(T)) + \langle T - WH ,\Lambda\rangle 
      + \tfrac{\rho}{2} \|T - WH\|_F^2.
\end{equation*} 
Since $T^{k+1}$ minimizes 
$L_\rho(W^{k+1}, H^{k+1}, T, \Lambda^k)$, we have 
\begin{align*}
0 &\in \partial[d(X,f(T^{k+1}))] 
    + \Lambda^k + \rho\,(T^{k+1} - W^{k+1}H^{k+1}) \\
  &= \partial[d(X,f(T^{k+1}))] + \Lambda^k + \rho \mathcal{R}^{k+1},
\end{align*}
where $\mathcal{R}^{k+1} := T^{k+1} - W^{k+1}H^{k+1}$ 
is the \emph{primal residual}.  
Rearranging gives 
\[
0 \in \partial[d(X,f(T^{k+1}))] + \Lambda^{k+1},
\]
which shows that $(T^{k+1}, \Lambda^{k+1})$ always satisfy \eqref{eq:dual_feasibility}.

Similarly, since $H^{k+1}$ minimizes 
$L_\rho(W^{k+1}, H, T^k, \Lambda^k)$,  
\begin{align*}
0 
&= -W^{k+1\top}\Lambda^k 
   - \rho W^{k+1\top}(T^k - W^{k+1}H^{k+1}) \\
&= W^{k+1\top}\big(\Lambda^k + \rho \mathcal{R}^{k+1} - \rho(T^{k+1} - T^k)\big) \\
&= W^{k+1\top}\big(\Lambda^{k+1} - \rho(T^{k+1} - T^k)\big).
\end{align*}
Hence $W^{k+1\top}\Lambda^{k+1}
    = \rho W^{k+1\top}(T^{k+1} - T^k)$. 
We define the \emph{dual residual} as
\[
 \mathcal{S}^{k+1} := \rho W^{k+1\top}(T^{k+1} - T^k).
\] 
Thus, $\mathcal{R}^{k+1}$ measures primal infeasibility, while $\mathcal{S}^{k+1}$ quantifies dual infeasibility from \eqref{eq:dual_feasibility2}.  
Finally, to balance primal and dual progress, we update $\rho$ at the end of each ADMM iteration using the following rule:
\begin{equation}\label{eq:adap_rho}
\rho_{k+1} :=
\begin{cases}
    \tau_{\mathrm{incr}} \cdot \rho_k 
        & \text{if } \| \mathcal{R}^k\|_F > \mu \| \mathcal{S}^k\|_F, \\[6pt]
    \rho_k / \tau_{\mathrm{decr}}
        & \text{if } \| \mathcal{S}^k\|_F > \mu \| \mathcal{R}^k\|_F, \\[6pt]
    \rho_k & \text{otherwise}, 
\end{cases}
\end{equation}
where 
\begin{itemize}
  \item $\mu > 1$ controls the relative tolerance,
  \item $\tau_{\mathrm{incr}} > 1$ and $\tau_{\mathrm{decr}} > 1$ scale $\rho$ up or down.
\end{itemize}

In our algorithm, we set $\mu = 10$ and 
$
\tau_{\mathrm{incr}} = \tau_{\mathrm{decr}} = 2,
$
which are also typical choices in the ADMM framework. 

Large values of $\rho$ impose a stronger penalty on violations of the primal feasibility condition, thereby driving the primal residuals toward zero. Conversely, by the definition of the dual residual $\mathcal{S}^k$, smaller values of $\rho$ reduce dual infeasibility, but simultaneously lessen the emphasis on primal feasibility. The adaptive update scheme \eqref{eq:adap_rho} balances these effects: $\rho$ is increased by a factor of $\tau_{\mathrm{incr}}$ whenever the primal residual is significantly larger than the dual residual, and decreased by a factor of $\tau_{\mathrm{decr}}$ whenever the dual residual dominates. In this way, the method dynamically adjusts $\rho$ to maintain a balance between primal and dual residuals. \revision{In Supplementary S3, we perform an ablation study comparing the adaptive strategy against fixed values 
of $\rho$ across multiple model/loss combinations. 
The results show that the adaptive strategy is robust across problem instances, consistently matching or outperforming the best fixed-$\rho$  configurations without requiring manual tuning of~$\rho$.}

\subsection{Handling missing data with ADMM}

Our proposed ADMM framework naturally handles missing data, which can be used for matrix completion, where the goal is to predict the value of missing entries. The updates for the factor matrices \(W\), \(H\), and the Lagrange multiplier \(\Lambda\) remain unchanged. The only modification arises in the update of the variable \(T\), which must treat observed and unobserved entries separately.   
Let \(\Omega\) denote the set of observed entries in \(X\), that is, if \((i,j) \in \Omega\), the entry \(X_{ij}\) is available, while for \((i,j) \notin \Omega\), the entry is missing. 
The update of \(T\) is obtained by solving
\begin{equation*}%\label{eq:matrix_completion_T_update}
    \min_{T} \sum_{(i,j) \in \Omega} d\big(X_{ij}, f(T_{ij})\big) 
    + \langle T - WH, \Lambda \rangle 
    + \tfrac{\rho}{2} \| T - WH \|_F^2. 
\end{equation*}
For indices corresponding to observed data, i.e., \((i,j) \in \Omega\), the updates for \(T_{ij}\) are identical to the ones described previously, with model-specific forms summarized in Table~\ref{tab:Tupdates}.
For the missing entries, \((i,j) \notin \Omega\), the update reduces to solving 
    $\min_{T_{ij}} \; \langle T - WH, \Lambda \rangle 
    + \tfrac{\rho}{2} \| T - WH \|_F^2$,
which yields the closed-form expression 
\[
    T_{ij} = \frac{\rho \,(WH)_{ij} - \Lambda_{ij}}{\rho}, 
    \quad (i,j) \notin \Omega.
\] 
This highlights the strength of the ADMM approach: it can easily adapt to different objectives, making it also a flexible tool for matrix completion. \revision{This will be illustrated in Section~\ref{sec:matrixcompletion}.}

\subsection{Convergence of Algorithm~\ref{alg:admm_nmd}} 
\label{sec:convergence_discussion}

\revisionSP{Existing convergence results for nonconvex
ADMM~\cite{hong2016convergence, wang2019global} do not apply directly to
our setting, because the coupling constraint $T = WH$ is bilinear rather
than linear. However, since $T$ appears in this constraint with the
identity as coefficient, and since the subproblems in $W$ and $H$ are
strongly convex because of the ridge regularization, convergence can be
established under two assumptions: (i)~the function
$T \mapsto d(X, f(T))$ is continuously differentiable with a Lipschitz
continuous gradient, and (ii)~the penalty parameter $\rho$ is  kept fixed,
and large enough, after a finite number of iterations. Under these
assumptions, we prove in Section~S4 of the supplementary material that the
augmented Lagrangian decreases at every iteration, that the iterates remain
bounded, that the primal residual $\|T^k - W^kH^k\|_F$ converges to zero,
and that every limit point is a stationary point.

Assumption~(i) is not satisfied by our models in their original
form, since the ReLU, MinMax and Modulus functions and the $\ell_1$ norm
are not differentiable, and the KL divergence is not Lipschitz smooth near
zero. However, it holds for smooth approximations of these functions, such
as the softplus function $\frac{1}{\beta}\log(1+e^{\beta t})$ that approximates 
the ReLU arbitrarily well as $\beta$ increases. 
Extending the analysis to the nonsmooth models and to the adaptive update
of $\rho$ is a challenging direction of research.} 

%% file: 6-experiments.tex
\section{Numerical Experiments}
\label{sec:experiments}
In this section, we present a series of numerical experiments on synthetic and real datasets under various noise conditions. These experiments are designed to illustrate the flexibility of the proposed ADMM-based algorithm and its ability to adapt to different data characteristics as well as to the type  of noise.  

\revision{All algorithms are implemented in MATLAB R2024a, with 
the exception of the PyTorch/Adam baseline 
(Section~\ref{sec:relu_comparison}), which is 
implemented in Python using PyTorch.
All experiments are executed on CPU on a MacBook Pro 
equipped with an Apple M2 processor and 8 GB of RAM.} The code implementing ADMM for the 12 models, along with all the experiments, is available from   
{\color{blue}\url{https://gitlab.com/Atharva05/admm-for-nmd}}. 
\revision{We note that the $T$-update is implemented using 
vectorized MATLAB operations for all model/loss 
combinations except CSF + Frobenius, which requires 
element-wise root-finding via Cardano's formula. In all 
cases, the dominant computational cost per iteration is 
the $W$ and $H$-updates, which involve matrix 
multiplications requiring $\mathcal{O}(mnr)$ operations.}

\paragraph*{Initialization of ADMM (Algorithm~\ref{alg:admm_nmd})} By default, the auxiliary variable $T$ is set to $T = X$, except for the CSF model where $T = X^{.1/2}$.  
Given the rank-$r$ truncated SVD  of $X$, $X \approx U \Sigma V^\top$, the factor matrices are initialized 
using  
$W = U \sqrt{\Sigma}$ and $H = \sqrt{\Sigma} V^\top$. 
The Lagrange multiplier is initialized at zero, i.e., $\Lambda = 0$. 
\revision{Note that cheaper initializations could be used, e.g., random initialization, or using randomized SVD algorithms; see, e.g.,~\cite{tropp2017practical}.}

\revisionSP{
\paragraph*{Choice of the rank} 
The choice of the rank $r$ is a classical model-selection problem and generally depends on the application at hand. In some applications, prior knowledge may provide a natural value or range for $r$, e.g., the number of sources in an audio sequence or the number of materials in a hyperspectral image.  
When such information is unavailable, a standard approach is to use cross-validation. More precisely, one can randomly hide a subset of the observed entries, compute the low-rank approximation using the remaining entries, and evaluate how accurately the resulting model predicts the held-out entries. Repeating this procedure for several data splits and different candidate values of $r$ provides an estimate of the out-of-sample performance for each rank. The rank yielding the best validation performance can then be selected. 

In the numerical experiments on real data, we do not explore this important issue as it is out of the scope of the paper. We chose a value for the rank that leads to reasonable reconstruction errors and good visual results for the image examples.  
}

\subsection{Synthetic data} 

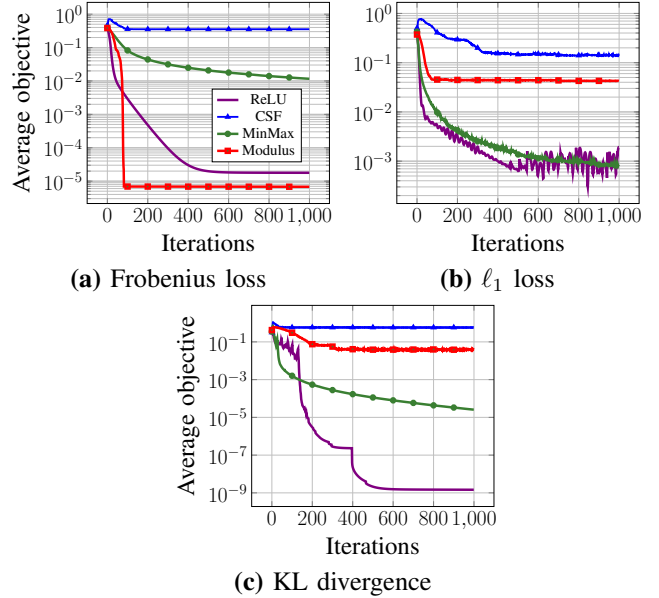
\begin{figure}[!htbp]
\centering

\begin{tikzpicture}
  \tikzset{lgd/.style={font=\scriptsize, anchor=west, inner sep=1pt}}
  \draw[violet, line width=1.3pt] (0,0) -- (0.42,0);
  \node[lgd] at (0.47,0) {ReLU};

  \draw[blue, line width=1.1pt] (1.30,0) -- (1.72,0);
  \node[blue] at (1.51,0) {\pgfuseplotmark{triangle*}};
  \node[lgd] at (1.77,0) {CSF};

  \draw[OliveGreen, line width=1.3pt] (2.52,0) -- (2.94,0);
  \node[OliveGreen] at (2.73,0) {\pgfuseplotmark{*}};
  \node[lgd] at (2.99,0) {MinMax};

  \draw[red, line width=1.3pt] (4.15,0) -- (4.57,0);
  \node[red] at (4.36,0) {\pgfuseplotmark{square*}};
  \node[lgd] at (4.62,0) {Modulus};
\end{tikzpicture}

\vspace{0.4em}

\begin{minipage}[t]{0.33\columnwidth}
\centering
\begin{tikzpicture}
  \begin{axis}[
    ymode=log,
    width=0.98\linewidth, height=0.92\linewidth,
    grid=both,
    grid style={line width=0.2pt, draw=gray!25},
    major grid style={line width=0.3pt, draw=gray!40},
    xlabel={Iterations},
    xlabel style={font=\scriptsize, yshift=2pt},
    ylabel style={font=\scriptsize, yshift=-4pt},
    tick label style={font=\tiny},
    xtick={0,500,1000},
    ytick distance=1e3,
    log basis y=10,
    scaled x ticks=false,
    ylabel={Average objective},
  ]
    \addplot[line width=1.3pt, violet]
      table [x expr=\coordindex, y=y1, col sep=space] {froresults.txt};
    \addplot[line width=1.1pt, blue, mark=triangle*,
      mark options={scale=0.55}, mark repeat=140]
      table [x expr=\coordindex, y=y2, col sep=space] {froresults.txt};
    \addplot[line width=1.3pt, OliveGreen, mark=*,
      mark options={scale=0.55}, mark repeat=140]
      table [x expr=\coordindex, y=y3, col sep=space] {froresults.txt};
    \addplot[line width=1.3pt, red, mark=square*,
      mark options={scale=0.55}, mark repeat=140]
      table [x expr=\coordindex, y=y4, col sep=space] {froresults.txt};
  \end{axis}
\end{tikzpicture}

\vspace{0.2em}
\textbf{(a)} Frobenius loss
\end{minipage}
\begin{minipage}[t]{0.32\columnwidth}
\centering
\begin{tikzpicture}
  \begin{axis}[
    ymode=log,
    width=0.98\linewidth, height=0.92\linewidth,
    grid=both,
    grid style={line width=0.2pt, draw=gray!25},
    major grid style={line width=0.3pt, draw=gray!40},
    xlabel={Iterations},
    xlabel style={font=\scriptsize, yshift=2pt},
    ylabel style={font=\scriptsize, yshift=-4pt},
    tick label style={font=\tiny},
    xtick={0,500,1000},
    ytick distance=1e3,
    log basis y=10,
    scaled x ticks=false,
  ]
    \addplot[line width=1.3pt, violet]
      table [x expr=\coordindex, y=y1, col sep=space] {L1results.txt};
    \addplot[line width=1.1pt, blue, mark=triangle*,
      mark options={scale=0.55}, mark repeat=140]
      table [x expr=\coordindex, y=y2, col sep=space] {L1results.txt};
    \addplot[line width=1.3pt, OliveGreen, mark=*,
      mark options={scale=0.55}, mark repeat=140]
      table [x expr=\coordindex, y=y3, col sep=space] {L1results.txt};
    \addplot[line width=1.3pt, red, mark=square*,
      mark options={scale=0.55}, mark repeat=140]
      table [x expr=\coordindex, y=y4, col sep=space] {L1results.txt};
  \end{axis}
\end{tikzpicture}

\vspace{0.2em}
\textbf{(b)} $\ell_1$ loss
\end{minipage}
\begin{minipage}[t]{0.32\columnwidth}
\centering
\begin{tikzpicture}
  \begin{axis}[
    ymode=log,
    width=0.98\linewidth, height=0.92\linewidth,
    grid=both,
    grid style={line width=0.2pt, draw=gray!25},
    major grid style={line width=0.3pt, draw=gray!40},
    xlabel={Iterations},
    xlabel style={font=\scriptsize, yshift=2pt},
    ylabel style={font=\scriptsize, yshift=-4pt},
    tick label style={font=\tiny},
    xtick={0,500,1000},
    ytick distance=1e3,
    log basis y=10,
    scaled x ticks=false,
  ]
    \addplot[line width=1.3pt, violet]
      table [x expr=\coordindex, y=y1, col sep=space] {KLresults.txt};
    \addplot[line width=1.1pt, blue, mark=triangle*,
      mark options={scale=0.55}, mark repeat=140]
      table [x expr=\coordindex, y=y2, col sep=space] {KLresults.txt};
    \addplot[line width=1.3pt, OliveGreen, mark=*,
      mark options={scale=0.55}, mark repeat=140]
      table [x expr=\coordindex, y=y3, col sep=space] {KLresults.txt};
    \addplot[line width=1.3pt, red, mark=square*,
      mark options={scale=0.55}, mark repeat=140]
      table [x expr=\coordindex, y=y4, col sep=space] {KLresults.txt};
  \end{axis}
\end{tikzpicture}

\vspace{0.2em}
\textbf{(c)} KL divergence
\end{minipage}

\caption{Average objective value as a function of the iteration number
on synthetic data $X\in\mathbb{R}^{100\times 80}$ generated from
rank-$5$ NMD models. Curves report means over 10 runs for each
nonlinearity under (a) Frobenius, (b) $\ell_1$, and (c) KL losses.}
\label{fig:synthetic-convergence}
\end{figure}

Let us first illustrate the behavior of the proposed ADMM framework across all 12 model configurations (3 loss functions, 4 nonlinear functions) on synthetic data. 
We generate the data matrices $X \in \mathbb{R}^{m \times n}$ with $m = 100$ and $n = 80$ by sampling the factor matrices $W \in \mathbb{R}^{m \times r}$ and $H \in \mathbb{R}^{r \times n}$ from a standard Gaussian, with the MATLAB command \texttt{randn}. We fix $r = 5$ and construct the data matrix as
$
X = f(WH),
$
where $f$ denotes a nonlinear function applied entrywise. Specifically, we consider the ReLU, CSF, MinMax, and Modulus nonlinearities given by $f(t) = \max(0,t)$, $f(t) = t^{2}$, $f(t) = \min(b,\max(a,t))$, and $f(t) = |t|$, respectively. 
For each combination, we perform a rank-5 factorization and run the ADMM algorithm for $10^3$ iterations.
Figure~\ref{fig:synthetic-convergence} reports the evolution of the average objective value over 10 runs. 

When the Frobenius norm or the $\ell_1$ norm is used, the objective corresponds to a relative reconstruction error between the data matrix $X$ and its nonlinear approximation $\hat{X} = f(WH)$. Specifically, the Frobenius objective is computed as $\|X - \hat{X}\|_F / \|X\|_F$, while the $\ell_1$ objective is given by $\|X - \hat{X}\|_1 / \|X\|_1$. For the Kullback--Leibler (KL) divergence, we define the objective as a normalized relative error $\mathrm{KL}(X,\hat{X}) / \mathrm{KL}(X,X_{\mathrm{mean}})$, where $X_{\mathrm{mean}}$ is a constant matrix with all entries equal to $\frac{1}{mn}\sum_{i=1}^{m}\sum_{j=1}^{n} X_{ij}$.

In all cases, we observe that ADMM converges well, except for the $\ell_1$ loss, for which it is more unstable; the reason being probably that the $\ell_1$ loss is not differentiable. 
Interestingly, for the nonlinearities CSF and Modulus (that is, $f(t) = t^2$ and $f(t) = |t|$), ADMM does not converge to small relative errors. 
The reason is that these nonlinearities are much more complex. In fact, with ReLU and MinMax, in the absence of noise, that is, $X = f(WH)$, we know a priori the range of values of $WH$. For example, for ReLU, we know that $(WH)_{ij} > 0$ when $X_{ij} > 0$ and $(WH)_{ij} \leq 0$ when $X_{ij} = 0$~\cite{seraghiti2023accelerated}. 
With CSF and Modulus, the algorithm has to find the right sign for the entries of $WH$, which is a combinatorial problem, making it significantly more challenging to solve. If the signs \revision{were}  given, then the problem would boil down to a standard matrix factorization problem, as already noted in~\cite{lefebvre2024component} for CSF. We further discuss this issue in Supplementary S2 where we show that in simpler scenarios (namely, smaller matrices, or when $W$ and $H$ are nonnegative), ADMM finds solutions with small relative errors.

\subsection{MNIST with Salt and Pepper Noise}

This experiment illustrates the flexibility of the proposed ADMM\revision{-}based algorithm and its ability to adapt to non-Gaussian noise scenarios. We use the MNIST handwritten digits dataset, which contains 60,000 grayscale images of size $28 \times 28$ pixels \cite{lecun2002gradient}. Each column of the data matrix $X$ corresponds to a vectorized handwritten digit. To ensure that all pixel intensities lie within the interval $[0,1]$, all entries of $X$ are divided by the maximum value of $X$. For our experiment, we randomly select 50 images per digit, resulting in a total of 500 images. A low-rank factorization with $\text{rank}=32$ is performed on the dataset of size $500 \times 784$.

Since ReLU-NMD has been shown to be particularly effective for sparse, nonnegative data \cite{seraghiti2023accelerated}, we adopt this model in our experiments.
Moreover, as the data values are bounded in $[0,1]$, we also consider the MinMax model, which is designed for such constrained ranges. To introduce corruption, we add \textit{salt and pepper noise} using the MATLAB built-in function \texttt{imnoise}.
Specifically, this function first assigns to each pixel a random probability drawn from a standard uniform distribution on the open interval $(0,1)$. Pixels with probability in $(0, d/2)$ are set to $0$ (``pepper''), while those in $[d/2, d)$ are set to the maximum intensity value (``salt''). The remaining fraction $(1-d)$ of pixels is left unchanged.
Hence, the noise density $d$ controls the proportion of entries replaced by extreme-valued outliers.
\begin{figure}[htb]
  \centering
  \includegraphics[width=0.8\textwidth]{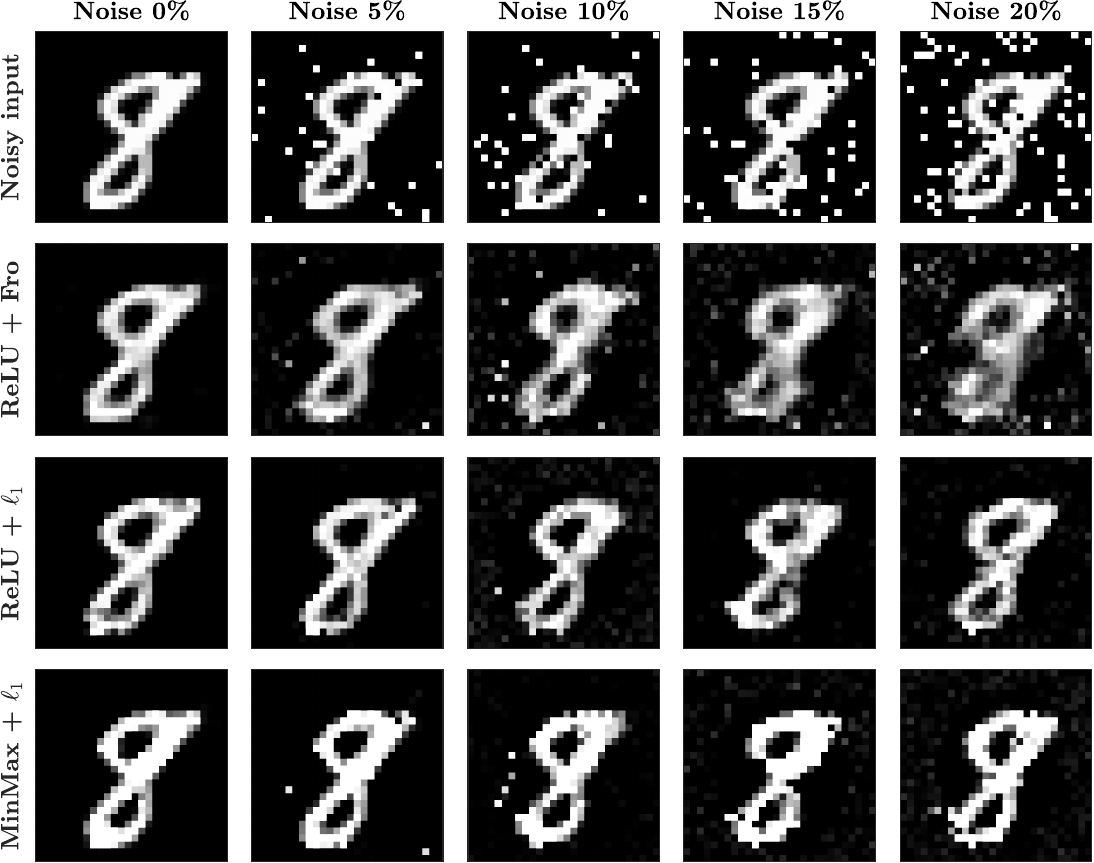}
  \caption{MNIST reconstruction of a digit under salt–and–pepper noise.  
  Rows: noisy input, ReLU\,+\,Frobenius, ReLU\,+\,$\ell_1$, MinMax\,+\,$\ell_1$.  
  Columns: noise density $d\in\{0\%,5\%,10\%,15\%,20\%\}$.} 
  \label{fig:mnist-snp}
\end{figure}

The purpose of this experiment is to show that the Frobenius norm is not well suited for handling non-Gaussian noise such as salt and pepper noise. 
In such cases, the ADMM framework can be used to employ more robust loss functions, such as the $\ell_1$-norm, which provides improved resilience to sparse, large-magnitude corruptions. Accordingly, we evaluate and compare the performance of the following models: ReLU+Frobenius, ReLU+$\ell_1$, and MinMax+$\ell_1$. Reconstruction results are reported for noise densities $d = \{0\%, 5\%, 10\%, 15\%, 20\%\}$. For each configuration and noise level, the algorithm is executed for 30 seconds.

Fig.~\ref{fig:mnist-snp} illustrates the reconstruction of an MNIST digit under different nonlinear models, loss functions, and noise levels. 
The Frobenius norm deteriorates significantly in the presence of outlier noise, whereas the $\ell_1$ norm exhibits more robustness to such corruptions, resulting in substantially improved reconstructions. Furthermore, a comparison between  ReLU+$\ell_1$ and MinMax+$\ell_1$ reveals distinct behaviors. ReLU+$\ell_1$ produces visually cleaner images but tends to slightly underestimate the pixel intensities of the digits. %leading to reconstructions that are less faithful to the original image. 
In contrast, the MinMax+$\ell_1$ model, by enforcing bounded optimization within the interval $[0,1]$, better preserves the intensity range of the digits. %and yields reconstructions that are closer to the true, noise-free images.

 \begin{table}[htb]
\centering
\caption{Relative reconstruction error with respect to the clean ground truth,
defined as
$\|X_{\mathrm{clean}} - X_{\mathrm{recon}}\|_{F} \,/\, \|X_{\mathrm{clean}}\|_{F}$,
for different noise levels and models.}
\label{tab:relative_error_noise}
\footnotesize
\setlength{\tabcolsep}{5pt}
\begin{tabular}{c c c c c}
\toprule
\textbf{Noise (\%)} 
& \textbf{Noisy vs GT}
& \textbf{ReLU + Fro}
& \textbf{ReLU + $\ell_1$}
& \textbf{MinMax + $\ell_1$} \\
\midrule
0  & 0      & 0.15786 & 0.21586 & \textbf{0.15120} \\
5  & 0.47086 & 0.34333 & 0.\textbf{28557} & 0.29985 \\
10 & 0.66348 & 0.41888 & 0.\textbf{34510} & 0.36894 \\
15 & 0.81196 & 0.48343 & 0.\textbf{40343} & 0.45480 \\
20 & 0.93860 & 0.57201 & 0.\textbf{46286} & 0.51534 \\
\bottomrule
\end{tabular}
\end{table}

Table~\ref{tab:relative_error_noise} provides the relative errors that are consistent with the visual comparisons shown in Fig.~\ref{fig:mnist-snp}. In the noiseless case, the MinMax+$\ell_1$ model achieves the lowest reconstruction error, benefiting from the explicit bounded pixel values. However, as the noise level increases, ReLU+$\ell_1$  consistently yields lower errors than MinMax+$\ell_1$. This indicates that while MinMax effectively preserves intensity bounds for clean data, it has a harder time finding good factorizations in noisier settings. Since applying the function \revision{$\min(1,\cdot)$} to any solution can only reduce the error, MinMax+$\ell_1$ should, in theory, always lead to lower reconstruction errors than ReLU+$\ell_1$. 
The reason this is not the case is that optimizing directly the MinMax objective is computationally harder and introduces  local minima in which MinMax+$\ell_1$ gets stuck. 
In contrast, ReLU+$\ell_1$ does not clip  \revision{the approximation within the interval $[0,1]$ during the optimization process, only within $[0,+\infty)$. 
This allows the convergence to better solutions in the noisy scenarios. Hence a direction of further research would be to design a homotopy method for MinMax, e.g., it would start with a larger interval for the MinMax model and then reduce it progressively during the iterative process until it reaches the sought interval $[a,b]$.}  %smaller intervals at each iteration, that is, $[a_1,b_1] \supseteq [a_2,b_2] \supseteq \dots \supseteq [a_k,b_k]$ where $[a_k,b_k]$ is the interval considered at iteration $k$, and  $[a_k,b_k]= [a,b]$ for all $k \geq K$ for some hyperparameter $K$.}   

\subsection{Matrix Completion on the CBCL Dataset} \label{sec:matrixcompletion}

\revision{We now evaluate the ability of our proposed ADMM framework to perform matrix completion. The experiments are conducted on the CBCL face dataset, which contains $2429$ grayscale images of size $19 \times 19$ pixels. All entries of $X$ are normalized by the maximum pixel value so that $X \in [0,1]$.

To simulate missing data, we retain $p\%$ of the entries of $X$ as observed. Among these observed entries, $80\%$ are used for training and the remaining $20\%$ are withheld for evaluation. During optimization, the test entries are set to zero, and the reconstruction quality is assessed using the root mean squared error (RMSE) computed on the test set. This setup measures the ability of the algorithms to accurately predict missing values. To account for variability due to the random selection of observed entries, we repeat each experiment over 10 independent random masks and report the mean and standard deviation of the test RMSE.

We compare four methods: 
\begin{enumerate}

    \item Our proposed ADMM framework with the MinMax model under the Frobenius loss. Since all pixel values lie in $[0,1]$, the MinMax nonlinearity $f(t) = \min(1, \max(0, t))$ is a natural choice. ADMM is run for 100 iterations. 

    \item Weighted low-rank approximation (WLRA) from \cite{gillis2020nonnegative}, which solves
\[
\min_{W \in \mathbb{R}^{m \times r},\; H \in \mathbb{R}^{n \times r}}
\| X - WH^\top \|_{M}^{2}
+\lambda \bigl( \|W\|_{F}^{2} + \|H\|_{F}^{2} \bigr),
\]
where the weighted Frobenius norm is defined as $\| X - WH^\top \|_{M}^{2} = \sum_{i,j} M_{ij}( X_{ij} - (WH^\top)_{ij})^{2}$. 

\item LMaFit \cite{wen2012solving}, an efficient alternating least-squares approach for low-rank matrix completion that minimizes $\|X_\Omega - (WH)_\Omega\|_F^2$ over factors $W \in \mathbb{R}^{m \times r}$ and $H \in \mathbb{R}^{r \times n}$, using a successive over-relaxation scheme. We run LMaFit for 100 iterations. \footnote{LMaFit code: 
\url{https://github.com/optsuite/LMaFit}}

\item Riemannian trust-region method for matrix completion (RTRMC)~\cite{boumal2011rtrmc}, implemented using the Manopt toolbox~\cite{boumal2014manopt}. This approach parametrizes the set of rank-$r$ matrices as a smooth Riemannian manifold in which the approximation $X_r$ belongs, and minimizes $\sum_{(i,j) \in \Omega} (X - X_r)_{ij}^2$ using a second-order trust-region solver.\footnote{Manopt toolbox: \url{https://www.manopt.org/}} 
\end{enumerate}  
All methods are run for 100 iterations and use the same rank $r=5$ and the same train/test split for each run to ensure a fair comparison. %The RMSE values on the test set for different missing-data ratios are reported in Table~\ref{table:ADMMvsWLRA}.
\begin{table}[htp]
\centering
\setlength{\tabcolsep}{4pt}
\caption{\revision{Matrix completion on the CBCL dataset ($r{=}5$).
Test RMSE (mean $\pm$ std over 10 random masks). ADMM is the proposed
method; the best result in each row is in bold.}}
\label{table:ADMMvsWLRA}
\begin{tabular}{@{}lcccc@{}}
\toprule
\textbf{Missing} & \textbf{ADMM} & \textbf{WLRA} & \textbf{LMaFit} & \textbf{RTRMC} \\
\midrule
0\%  & $\mathbf{0.1024 \pm 0.0002}$ & $0.1459 \pm 0.0003$ & $0.1025 \pm 0.0002$ & $0.1026 \pm 0.0002$ \\
5\%  & $\mathbf{0.1025 \pm 0.0003}$ & $0.1459 \pm 0.0003$ & $0.1026 \pm 0.0003$ & $0.1026 \pm 0.0003$ \\
10\% & $\mathbf{0.1025 \pm 0.0003}$ & $0.1459 \pm 0.0003$ & $0.1026 \pm 0.0003$ & $0.1026 \pm 0.0003$ \\
20\% & $\mathbf{0.1026 \pm 0.0001}$ & $0.1460 \pm 0.0002$ & $0.1027 \pm 0.0001$ & $0.1027 \pm 0.0001$ \\
50\% & $\mathbf{0.1037 \pm 0.0003}$ & $0.1462 \pm 0.0003$ & $\mathbf{0.1037 \pm 0.0003}$ & $\mathbf{0.1037 \pm 0.0003}$ \\
80\% & $0.1085 \pm 0.0008$ & $0.1471 \pm 0.0009$ & $0.1082 \pm 0.0007$ & $\mathbf{0.1081 \pm 0.0007}$ \\
\bottomrule
\end{tabular}
\end{table}

Table~\ref{table:ADMMvsWLRA} reports the RMSE on the test set for different ratios of missing data. It shows that ADMM, LMaFit, and RTRMC all achieve comparable test RMSE across all levels of missing data, consistently outperforming WLRA\footnote{The reason could be that WLRA uses a specific initialization: it constructs an  initial solution in a greedy fashion, one rank-one factor at a time,  which could lead to worse solutions in this experiment.}. 
For missing ratios up to 50\%, ADMM holds a marginal advantage. At 80\% missing data, all three degrade, with RTRMC and LMaFit achieving a slightly lower RMSE of $0.1081$ and $0.1082$, resp., compared to $0.1085$ for ADMM, a negligible difference. %In contrast, WLRA yields substantially higher RMSE values across all settings (${\sim}0.146$).   %indicating that its linear model is insufficient for this data.

These results illustrate that the proposed ADMM framework, despite being designed as a general-purpose solver for NMDs, achieves accuracy on par with dedicated state-of-the-art matrix completion algorithms. Crucially, unlike LMaFit and RTRMC, which are restricted to linear low-rank models, our ADMM algorithm supports nonlinear  models and diverse loss functions, providing a unified framework that extends naturally to the matrix completion setting.
}

\subsection{Robustness to Poisson noise}

We use the standard low-rank benchmark image known as the \emph{MIT logo} which has rank 4, 
and add Poisson noise using the MATLAB function \texttt{poissrnd}. The image is a $250 \times 482$ grayscale matrix with pixel values in the interval $[0.5,1]$. Because the data is bounded, the MinMax nonlinearity is well suited for this task. In addition, Poisson noise is more naturally handled by the KL divergence, so this loss function is expected to perform better.

We compute a rank-4 factorization of the noisy image and compare the reconstruction quality using different loss functions (Frobenius, KL, and $\ell_1$) and different nonlinearities (such as ReLU). For all experiments, the MinMax bounds are fixed to $[0.5,1]$, and each algorithm is run for 10 seconds.

\begin{figure}[ht!]
\centering
\begin{tabular}{cc}

\includegraphics[width=0.4\textwidth]{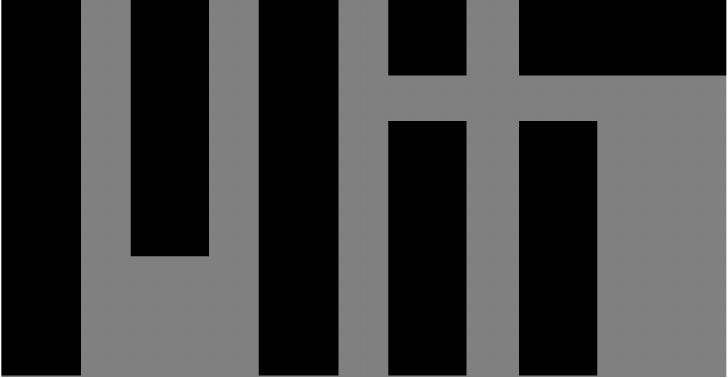} &
\includegraphics[width=0.4\textwidth]{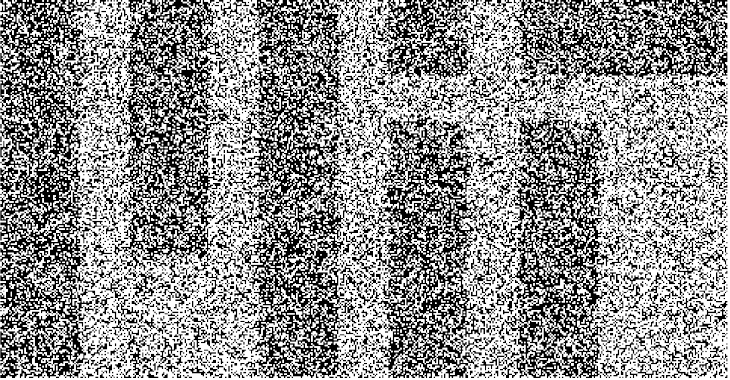} \\[-0.2em]
\small (a) Clean image &
\small (b) Noisy image \\[1.0em]

\includegraphics[width=0.4\textwidth]{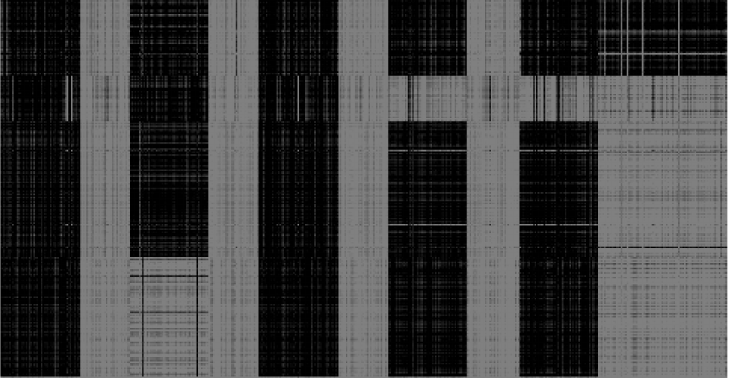} &
\includegraphics[width=0.4\textwidth]{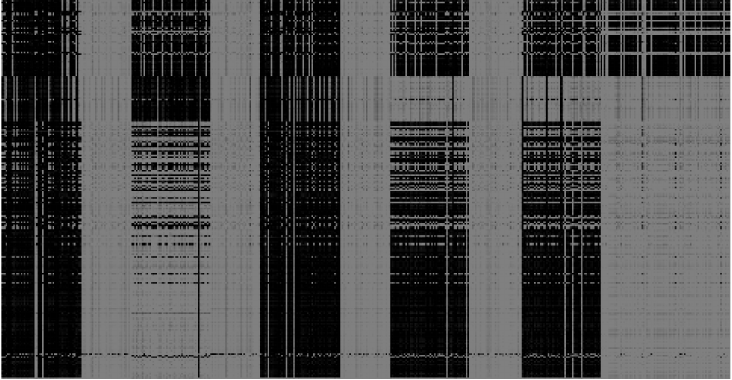} \\[-0.2em]
\small (c) MinMax + KL &
\small (d) MinMax + Frobenius \\[1.0em]

\includegraphics[width=0.4\textwidth]{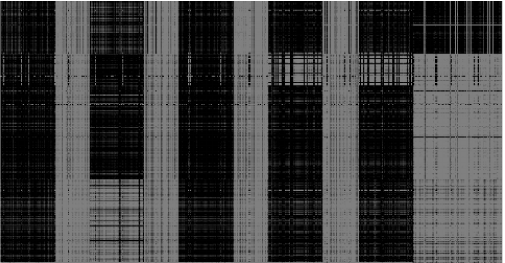} &
\includegraphics[width=0.4\textwidth]{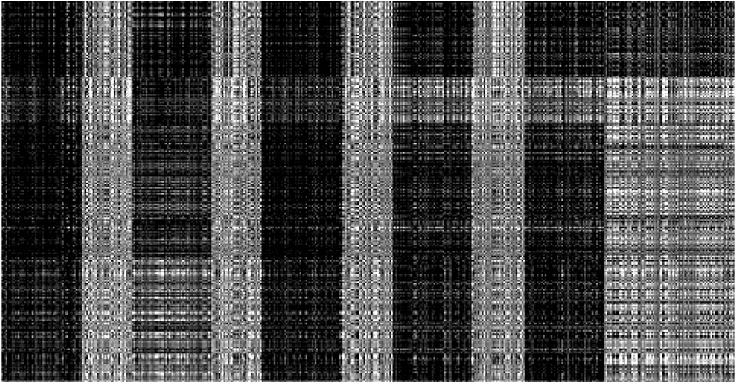} \\[-0.2em]
\small (e) MinMax + $\ell_1$ &
\small (f) ReLU + KL \\

\end{tabular}

\caption{Reconstruction results for the Poisson-corrupted MIT logo image under different nonlinear models and loss functions.}
\label{fig:poisson_grid}
\end{figure}

\begin{table}[htp]
\centering
\caption{Relative error 
$\|X_{\mathrm{clean}} - X_{\mathrm{recon}}\|_F / \|X_{\mathrm{clean}}\|_F$
for different nonlinearities and loss functions.}
\label{tab:relative_error}
\footnotesize
\setlength{\tabcolsep}{12pt}  % widen columns
\begin{tabular}{l c}
\toprule
\textbf{Model + Loss Function} 
& \textbf{Relative Error (\%)} \\
\midrule
MinMax + KL       &\textbf{ 9.89} \\
MinMax + Frobenius & 19.94 \\
MinMax + $\ell_1$ & 12.84 \\
ReLU + KL         & 29.48 \\
\bottomrule
\end{tabular}
\end{table}

The reconstruction results in Fig.~\ref{fig:poisson_grid} show that the combination of the MinMax nonlinearity with the KL divergence yields the most accurate recovery of the Poisson-corrupted MIT logo image, effectively removing  the noise significantly better than the other variants. In contrast, the MinMax model paired with either the Frobenius or $\ell_1$ loss performs noticeably worse, which is expected since these losses correspond to Gaussian and Laplacian noise models and are therefore not well suited to Poisson noise. The ReLU nonlinearity combined with the KL divergence also leads to poor reconstruction quality. This behavior is consistent with the fact that ReLU-based NMD performs best on sparse data, a condition not satisfied in this experiment, and is further supported by the quantitative errors reported in Table~\ref{tab:relative_error}. Overall, MinMax provides a more suitable nonlinearity than ReLU for this task, as it respects the bounded nature of the data and does not rely on sparsity assumptions.

\subsection{Comparison of ADMM with State-of-the-art for ReLU}
\label{sec:relu_comparison}
\revision{The purpose of this experiment is to compare the performance of the proposed ADMM algorithm with two baselines for solving ReLU-NMD with the Frobenius norm: 
(1)~the dedicated coordinate descent (CD) method from \cite{awari2024coordinate}, and (2)~the Adam optimizer~\cite{kingma2014adam} applied via PyTorch's automatic differentiation. CD is currently the only algorithm that directly addresses the original ReLU-NMD formulation, without reformulating it as a non-equivalent latent variable model. In contrast, several existing approaches rely on such reformulations; see \cite{saul2022nonlinear}, \cite{seraghiti2023accelerated}, and \cite{gillis2025extrapolated}. Since the proposed ADMM method also operates directly on the original ReLU-NMD problem, it provides a natural and fair basis for comparison with CD. Additionally, we include PyTorch/Adam as a representative generic first-order method to evaluate whether a problem-agnostic gradient-based approach can match the performance of structure-exploiting algorithms. 
The experiments are conducted on two datasets: (1)~the CBCL face dataset, which consists of 2429 grayscale facial images of size $19 \times 19$, and (2)~a subset of the MNIST handwritten digits dataset, containing 50 images per digit class (500 images total) of size $28 \times 28$. We perform a rank-10 factorization for CBCL and a rank-32 factorization for MNIST. All three algorithms are initialized with the same factor matrices $W$ and $H$, sampled from a Gaussian distribution and subsequently scaled according to the magnitude of the entries of the data matrix~$X$. To ensure a fair comparison, all methods are given the same time budget of 10 seconds per run.  
To account for sensitivity to initialization, we repeat each experiment over 10 independent random initializations and report the mean and standard deviation of the final relative reconstruction error.

Figure~\ref{fig:admmvscd} reports the results, and summary statistics are provided in Table~\ref{tab:admmvscd}. 

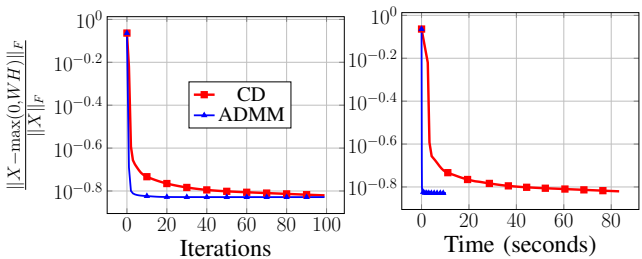
\begin{figure}[!htbp] 
\centering
\begin{minipage}[t]{0.45\columnwidth}
        \centering
        \begin{tikzpicture}[scale=0.78]
            \begin{axis}[
                ymode=log,
                xlabel={Time (seconds)},
                ylabel = Average objective,
                grid=both,
                legend style={at={(0.62, 0.78)}, anchor=north, legend columns=1, font=\large}, 
                tick label style={font=\large},
                xlabel style={font=\Large}, 
                ylabel style={font=\Large},
            ]

                \addplot[line width=2pt, color=OliveGreen, mark=*, mark repeat=20] table [x=x1, y=y1, col sep=space] {PyTorch_time_CBCL.txt};
                
                \addplot[line width=2pt, color=red, mark=square*, mark repeat=15] table [x=x1, y=y1, col sep=space] {ADMMvsCD_time_CBCL.txt};
    
                \addplot[line width=1.5pt, color=blue, mark=triangle*, mark repeat=30] table [x=x2, y=y2, col sep=space] {ADMMvsCD_time_CBCL.txt};

            \end{axis}
        \end{tikzpicture}

\textbf{(a)} CBCL ($r{=}10$)
\end{minipage}
\hfill 
\begin{minipage}[t]{0.45\columnwidth}
        \centering
        \begin{tikzpicture}[scale=0.78]
            \begin{axis}[
                ymode=log,
                xlabel={Time (seconds)},
                grid=both,
                legend style={at={(0.62, 0.94)}, anchor=north, legend columns=1, font=\large}, 
                tick label style={font=\large},
                xlabel style={font=\Large}, 
                ylabel style={font=\Large},
            ]

                \addplot[line width=2pt, color=OliveGreen, mark=*, mark repeat=15] table [x=x1, y=y1, col sep=space] {PyTorch_time_MNIST.txt};
                \addlegendentry{PyTorch/Adam}

                \addplot[line width=2pt, color=red, mark=square*, mark repeat=15] table [x=x1, y=y1, col sep=space] {ADMMvsCD_time_MNIST.txt};
                \addlegendentry{CD}
    
                \addplot[line width=1.5pt, color=blue, mark=triangle*, mark repeat=30] table [x=x2, y=y2, col sep=space] {ADMMvsCD_time_MNIST.txt};
                \addlegendentry{ADMM}

            \end{axis}
        \end{tikzpicture}

\textbf{(b)} MNIST ($r{=}32$)
\end{minipage}
\caption{\revision{Comparison of ADMM, CD, and PyTorch/Adam for 
ReLU-NMD on the CBCL dataset (left) and the MNIST 
dataset (right). The average relative error (over 10 runs) is shown as a function of computational time. All algorithms are given a time budget of 10 seconds and initialized from the same starting points.}} 
\label{fig:admmvscd}
\end{figure}

\begin{table}[htbp]
\centering
\small
\caption{\revision{Comparison of ADMM, CD, and PyTorch/Adam for ReLU-NMD on
the CBCL and MNIST datasets, with a time budget of 10 seconds. Results are
averaged over 10 random initializations. The last two columns give the
number of iterations performed within the time budget and the average time
per iteration.}}
\label{tab:admmvscd}
\begin{tabular}{llccc}
\toprule
\textbf{Dataset} & \textbf{Method} & \textbf{Final rel. error} &
\textbf{Iterations} & \textbf{Time/iter. (ms)} \\
\midrule
\multirow{3}{*}{CBCL ($r{=}10$)}
 & PyTorch/Adam ($\alpha = 10^{-1}$) & $0.1584 \pm 0.0048$          & 1863 & 5.37 \\
 & CD~\cite{awari2024coordinate}     & $0.2278 \pm 0.1320$          & 31 & 392.97 \\
 & ADMM (proposed)                   & $\mathbf{0.1484 \pm 0.0000}$ & 319 & 31.57 \\
\midrule
\multirow{3}{*}{MNIST ($r{=}32$)}
 & PyTorch/Adam ($\alpha = 10^{-2}$) & $0.1825 \pm 0.0034$          & 4728 & 2.12 \\
 & CD~\cite{awari2024coordinate}     & $0.2556 \pm 0.0065$          & 27 & 375.30 \\
 & ADMM (proposed)                   & $\mathbf{0.1768 \pm 0.0014}$ & 646 & 15.50 \\
\bottomrule
\end{tabular}
\end{table}

Figure~\ref{fig:admmvscd} and Table~\ref{tab:admmvscd} show that ADMM achieves the lowest final error on both datasets. On CBCL, ADMM converges to $0.1484 \pm 0.0000$ 
within the first second, while PyTorch/Adam reaches $0.1584 \pm 0.0048$ and CD stagnates at $0.2278 \pm 0.1320$ after the full 10-second budget. On 
MNIST, ADMM and PyTorch/Adam achieve comparable errors ($0.1768$ vs $0.1825$), both significantly outperforming CD ($0.2556$). 

The PyTorch/Adam results reported here 
correspond to the best learning rate selected from a 
sweep over 
$\alpha \in \{10^{-4}, 10^{-3}, 10^{-2}, 10^{-1}, 0.5\}$, with $\alpha = 10^{-1}$ performing best on CBCL and $\alpha = 10^{-2}$ on MNIST (note that $\alpha = 10^{-3}$ is the default recommended value~\cite{kingma2014adam}). 
ADMM requires no such tuning and achieves lower error with its 
default parameters. Moreover, ADMM exhibits near-zero variance across initializations (standard deviation of $0.0000$ on CBCL and $0.0014$ on MNIST), while both PyTorch/Adam and CD show higher sensitivity, with CD particularly unstable on CBCL (standard deviation of $0.1320$). 

We have also performed the same experiment with the ReLU nonlinearity and the KL divergence: PyTorch/Adam was not able to converge (it diverges towards very high relative errors, larger than 1000, for all learning rates and all initializations, for both datasets), probably because the gradient becomes unbounded near 0. 

\revisionSP{The last two columns of Table~\ref{tab:admmvscd} report the number of
iterations performed within the time budget and the average time per iteration. They show that the fixed time budget does not disadvantage CD. Although each iteration of ADMM solves linear systems, these only involve $r \times r$ matrices, at a cost of $O(r^3 + (m+n)r^2)$, so that one iteration of ADMM costs $O(mnr)$ operations, dominated by the matrix products. In comparison, one iteration of CD costs $O(mnr \max(\log m, \log n))$ operations~\cite{awari2024coordinate}, because each update of a single entry of $W$ or $H$ requires sorting breakpoints. In practice, the difference is even larger: one iteration of CD takes about $393$~ms against $31.6$~ms for ADMM on CBCL, and $375$~ms
against $15.5$~ms on MNIST. This is because ADMM relies on matrix--matrix products, which are computed very efficiently, whereas CD updates the entries one at a time. Within $10$ seconds, ADMM therefore performs more than ten times more iterations than CD on CBCL, and more than twenty times more on MNIST. PyTorch/Adam performs even more iterations, since each of them is a single gradient step; such a step is cheaper, but makes less progress than an iteration of ADMM or CD, which is why we compare the
methods on the error reached within the same time rather than on the number of iterations.}
}

\revisionSPJ{We also repeated this experiment with a time budget of $60$ seconds. ADMM still reaches the lowest error on both datasets, namely $0.1484$ on CBCL and $0.1636$ on MNIST, compared to $0.1487$ and $0.1798$ for Adam, and $0.1604$ and $0.1797$ for CD.} %On CBCL, ADMM converges within $2$ seconds, while Adam needs about $40$ seconds to reach a comparable error; on MNIST, the advantage of ADMM over Adam increases with time.}

\revisionSP{\subsection{Square-root rank of linear Euclidean distance matrices}
\label{sup:ledm}
 
In this experiment, we compare our proposed ADMM framework with the coordinate descent (CD) algorithm of~\cite{lefebvre2024component}, which is designed specifically for CSF with the Frobenius loss. We use a task for which CSF is useful: computing the square-root rank of a matrix.
 
The square root rank of a nonnegative matrix $X \in \mathbb{R}^{m \times n}$ is defined as follows. A Hadamard square root of $X$ is a matrix of the same size whose $(i,j)$-th entry is either $\sqrt{X_{ij}}$ or $-\sqrt{X_{ij}}$. The square root rank of $X$ is the smallest rank among all Hadamard square roots of $X$. Equivalently, it is the smallest $r$ such that $X = (WH)^{.2}$ for some $W \in \mathbb{R}^{m \times r}$ and
$H \in \mathbb{R}^{r \times n}$.

The square root rank is used to study how compactly a polytope can be described. Some polytopes need a very large number of linear inequalities to be described, but admit a much smaller description using semidefinite constraints. The size of the smallest such description is equal to the positive semidefinite rank of the slack matrix of the
polytope, and the square root rank of this matrix is an upper bound on its positive semidefinite rank~\cite{fawzi2015positive}. The square root rank therefore gives a bound on how compactly a polytope can be described. For example, it was shown in~\cite{lee2014square} that the square root rank of the correlation polytope grows exponentially with its dimension.

Computing the square root rank is difficult. A matrix with $k$ nonzero entries has $2^k$ Hadamard square roots, one for each choice of signs, so trying all of them is only possible for very small matrices. This cannot be avoided in general, since computing the square root rank is NP-hard~\cite{fawzi2015positive,gillis2026complexity}. Instead, one can solve CSF with a given rank $r$: if a solution with zero error is found, that is,
$X = (WH)^{.2}$, then $r$ is an upper bound on the square root rank of
$X$. Solving CSF for decreasing values of $r$ therefore gives an estimate of the square root rank.
 
We use the linear Euclidean distance matrices (LEDMs), defined by $X_{ij} = (i-j)^2$ for $i,j = 1,\dots,n$. Their square root rank is equal to $2$ for every $n\geq2$~\cite{fawzi2015positive}. Since the answer is known, these matrices are a good test: a method succeeds if it finds an exact factorization with $r = 2$.
 
\paragraph{\textbf{Setup}} We use the experimental setup of~\cite{lefebvre2024component}, so that our results are directly comparable with theirs. In particular, we consider the same matrix sizes, $n = 3, \dots, 10$, with $r = 2$. For each $n$, we run both methods $100$ times. In each run, both methods start from the same random
initial point, generated as in~\cite{lefebvre2024component}: the entries of $W$ and $H$ are drawn from the standard Gaussian distribution, and $W$ and $H$ are then scaled to best fit $X$. For CD, we use the version with extrapolation of~\cite{lefebvre2024component}, with our MATLAB implementation of the code of the authors. For ADMM, we use the CSF model with the Frobenius loss. A run is successful if the relative error $\|X - (WH)^{.2}\|_F / \|X\|_F$ falls below $10^{-4}$. Both methods use the same stopping rules: a run stops when the error has not decreased by at least $1\%$ over the last $10$ iterations, as in~\cite{lefebvre2024component}, or after $10^4$ iterations, or after $2$ seconds. %For this rule, we use the smallest error reached so far, because the error of ADMM does not decrease at every iteration.
 
\paragraph{\textbf{Results}}
Table~\ref{tab:ledm} reports the percentage of successful runs. The success rates of CD are close to those reported in~\cite{lefebvre2024component}. For $n = 5, \dots, 10$, ADMM succeeds more often than CD, often by a large margin: for example, for $n = 6$,
ADMM succeeds in $95$ runs against $49$ for CD, and for $n = 10$, in $68$ runs against $28$. For $n = 4$, the two methods are close, and for $n = 3$, CD succeeds more often.
These results show that ADMM, although designed as a general framework for NMDs, recovers the exact factorization of the LEDMs more often than an algorithm designed specifically for CSF, for most of the matrix sizes we tested.
 
\begin{table}[htbp]
\centering
\caption{Percentage of successful runs, out of $100$, on the $n \times n$ LEDMs with $r = 2$. Both methods start from the same initial points. The highest value of each column is in bold.}
\label{tab:ledm}
\begin{tabular}{lcccccccc}
\toprule
$n$ & 3 & 4 & 5 & 6 & 7 & 8 & 9 & 10 \\
\midrule
CD~\cite{lefebvre2024component} & \textbf{85} & 76 & 64 & 49 & 60 & 43 & 55 & 28 \\
ADMM & 61 & \textbf{80} & \textbf{97} & \textbf{95} & \textbf{84} & \textbf{82} & \textbf{69} & \textbf{68} \\
\bottomrule
\end{tabular}
\end{table}
} 

%% file: 7-conclusion.tex
\section{Conclusion}
\label{sec:conclusion}
Nonlinear matrix decompositions (NMDs) are low-rank models of growing interest, with applications such as low-dimensional embedding, probabilistic circuit modeling, and recommender systems. Despite their broad applicability, NMDs remain challenging to solve in practice, particularly when combined with loss functions beyond the standard Frobenius norm. In this work, we proposed a flexible ADMM-based framework for NMDs that accommodates a wide class of nonlinear models and loss functions within a unified optimization scheme. Importantly, the proposed framework is designed to seamlessly incorporate new nonlinear models as they emerge, ensuring long-term extensibility.
Our approach draws inspiration from the classical ADMM methodology in convex optimization and adapts it to the nonlinear and nonconvex setting of NMDs. To enhance practical performance, we introduced an adaptive strategy for dynamically updating the penalty parameter during the optimization process. The proposed framework naturally extends to settings with missing data, making it particularly well suited for matrix completion problems.
Through extensive numerical experiments, we showed the flexibility and effectiveness of the proposed ADMM framework across a range of nonlinear models, loss functions, and problem settings, including full matrix factorization and incomplete data scenarios. In particular, for ReLU-NMD, the proposed ADMM method significantly outperforms the current state-of-the-art coordinate descent algorithm in both convergence speed and computational efficiency. These results highlight the potential of ADMM as a powerful and versatile optimization tool for NMDs.

%% file: supplementary.tex
\clearpage
\setcounter{section}{0}
\setcounter{subsection}{0}
\setcounter{equation}{0}
\setcounter{figure}{0}
\setcounter{table}{0}
\setcounter{algocf}{0}
\setcounter{theorem}{0}
\setcounter{lemma}{0}
\setcounter{corollary}{0}
\setcounter{assumption}{0}
\renewcommand{\thesection}{S\arabic{section}}
\renewcommand{\thesubsection}{S\arabic{section}.\arabic{subsection}}
\renewcommand{\theequation}{S\arabic{equation}}
\renewcommand{\thefigure}{S\arabic{figure}}
\renewcommand{\thetable}{S\arabic{table}}
\renewcommand{\thealgocf}{S\arabic{algocf}}
\renewcommand{\thetheorem}{S\arabic{theorem}}
\renewcommand{\thelemma}{S\arabic{lemma}}
\renewcommand{\thecorollary}{S\arabic{corollary}}
\renewcommand{\theassumption}{S\arabic{assumption}}

\begin{center}
{\Large\bfseries Supplementary Material}\\[6pt]
{\large Alternating Direction Method of Multipliers for Nonlinear Matrix Decompositions}
\end{center}
\vspace{1em}

\noindent
This supplementary material accompanies the article
\emph{``Alternating Direction Method of Multipliers for Nonlinear Matrix
Decompositions''}. Section~\ref{sup:updateT} derives the closed-form
solution of the scalar $T$-subproblem for each of the twelve
model/loss combinations. Section~\ref{sup:hardness} discusses the
intrinsic hardness of the CSF and Modulus models. Section~\ref{sup:rho}
reports an ablation study on the penalty parameter~$\rho$. Section~\ref{sec:convergence} analyzes the convergence of ADMM for NMD. Sections,
equations, figures and tables are numbered with the prefix~S; all
references to the main article use its own numbering.

\vspace{1em}
\hrule
\vspace{1em}

\section{Updates of $T$}
\label{sup:updateT}

This appendix presents the updates for the variable $T$ within our ADMM framework; see Table~1. 
At the matrix level, the update of $T$ is given by  
\begin{equation}
\label{eq:T-update-matrix}
T^{k+1} \;\in\; \argmin_{T \in \mathbb{R}^{m \times n}}
\; d(X, f(T))
\;+\; \langle T, \Lambda^k \rangle
\;+\; \frac{\rho}{2}\|T - A^{k+1}\|_F^2,
\end{equation}
where \(A^{k+1} := W^{k+1}H^{k+1}\). The objective in \eqref{eq:T-update-matrix} is separable across entries of \(T\); hence \eqref{eq:T-update-matrix} decomposes into \(mn\) independent one-dimensional subproblems. To derive entrywise update rules, we subsequently focus on a single element of $T$. Specifically, for each $(i,j)$, we denote $x = X_{ij}$, $a = A^{k+1}_{ij}$, $\lambda = \Lambda^k_{ij}$, and $t = T_{ij}$. 

All updates are implemented in our MATLAB code; see {\color{blue}
\url{https://gitlab.com/Atharva05/admm-for-nmd}}.

\subsection{CSF with Frobenius loss} \label{app:csf_frob}

The scalar subproblem is
$\min_{t\in\mathbb{R}}\; 
g(t):=\tfrac{1}{2}(x - t^2)^2 + \lambda t + \tfrac{\rho}{2}(t-a)^2$. 
A necessary condition for a stationary point is \(g'(t)=0\). Direct differentiation gives
\[
g'(t) = 2t^3 + (\rho - 2x)\,t + (\lambda - \rho a) = 0.
\]
This yields the cubic equation 
\begin{equation*}
t^3 + p\,t + q = 0,
\quad\text{with}\quad
p = \frac{\rho - 2x}{2},\quad q = \frac{\lambda - \rho a}{2}.
\end{equation*}  
Hence, the problem reduces to finding the real roots of a cubic equation, which can be done efficiently using Cardano’s method or any robust 1D root-finding procedure. The root that minimizes the scalar objective is selected as the updated entry.

\subsection{MinMax with Frobenius norm} \label{app:minmax_frob}

 The scalar problem is
\[
\min_{t\in\mathbb{R}} \;\frac{1}{2}(x - \min(q, \max(p,t)))^2 \;+\; \lambda t \;+\;\frac{\rho}{2}(t-a)^2.
\]
This splits into three quadratic functions in $t$:  
\[
\begin{array}{cl}
g_1(t)  =  \frac{1}{2}(x-q)^2 + t\lambda + \frac{\rho}{2}(t-a)^2    & \text{ for }  t > q, \\
 g_2(t)  =  \frac{1}{2}(x-t)^2 + t\lambda + \frac{\rho}{2}(t-a)^2   &  \text{ for }  p \leq t \leq q, \\ 
g_3(t)  =   \frac{1}{2}(x-p)^2 + t\lambda + \frac{\rho}{2}(t-a)^2 & \text{ for }  t < p. 
\end{array}
\] 
The candidate minimizers are
\[
t_1 = \frac{s}{\rho}, 
\qquad t_2 = \frac{x+s}{\rho+1}, 
\qquad t_3 = t_1,
\]
together with the boundary values $p$ and $q$, and $s = \rho a - \lambda$.  

The update is determined by threshold conditions on the parameter \(s=\rho a - \lambda\).  
The cases are summarized as follows:  
\begin{itemize}
    \item[\textbf{(i)}] 
    If $t_1^* = t_3^* > q$ and $t_2^* > q$, 
        $s > \rho q$ 
        and  
        $s > q + \rho q - x$,
    then the minimizer is in this region, and 
        $t^{k+1} = t_1^*$. 

    \item[\textbf{(ii)}] 
    If  $t_1^* = t_3^* \in  (p,q)$ 
    and $t_2^* \in (p,q)$; 
        $\rho p < s < \rho q$ 
        and 
        $p+\rho p - x < s < q+\rho q - x$,
    then the minimizer lies strictly within the interval, and $t^{k+1} = t_2^*$. 

    \item[\textbf{(iii)}]  
    If $t_1^* = t_3^* < p$ and $t_2^* < p$
        $s < \rho p$ and 
        $s < p + \rho p - x$, 
    then the minimizer lies in this region, and 
        $t^{k+1} = t_3^*$.

    \item[\textbf{(iv)}] 
    If $t_1^* = t_3^* < p$ and $t_2^* \in (p,q)$, 
        $s < \rho p$, $p+\rho p - x < s < q+\rho q - x$,
    then both \(t_2\) and \(t_3\) are feasible.  
    The update is obtained by comparing their objectives: 
    $t^{k+1} = t_2^*$ if $g_2(t_2^*) \leq g_3(t_3^*)$, and $t^{k+1} =t_3^*$ otherwise. 

    \item[\textbf{(v)}]  
    If $t_1^* = t_3^* > q$ and $t_2^* \in (p,q)$, 
        $s > \rho q$, and 
         $p+\rho p - x < s < q+\rho q - x$,
    then both \(t_1\) and \(t_2\) are feasible.  
    The update is obtained by comparing their objectives: $t^{k+1} = t_1^*$ if $g_1(t_1^*) \leq g_2(t_2^*)$, and $t^{k+1} =t_2^*$ otherwise. 

    \item[\textbf{(vi)}] 
    If $t_1^* = t_3^* \in (p,q)$ and $t_2^* > q$, 
        $\rho p < s < \rho q$, and 
        $s > q+\rho q - x$,
    then the minimizer is attained at the boundary, so 
        $t^{k+1} = q$.

    \item[\textbf{(vii)}]
    If $t_1^* = t_3^* \in (p, q)$ and $t_2^* <p$, 
        $\rho p < s < \rho q$ and 
        $s < p+\rho p - x$,
    then the minimizer is attained at the boundary, so %we set
        $t^{k+1} = p$.
\end{itemize}

\subsection{Modulus with Frobenius norm}  \label{app:modulus_frob}

The scalar problem is 
$ 
\min_t\frac{1}{2}(x - |t|)^2 + \lambda t + \frac{\rho}{2}(t-a)^2$,  
which can be split into two quadratic functions:
\[
g(t) = \left\{ 
\begin{array}{cc}
  \frac{1}{2}(x - t)^2 + \lambda t + \frac{\rho}{2}(t-a)^2 
  & \text{ for } t > 0,\\
\frac{1}{2}(x + t)^2 + \lambda t + \frac{\rho}{2}(t-a)^2  
& \text{ for } t \le 0.  
\end{array} 
\right.
\]   
Their unconstrained minimizers are
\[
t_1^* = \frac{\rho a - \lambda + x}{\rho + 1}, \qquad
t_2^* = \frac{\rho a - \lambda - x}{\rho + 1}.
\]   
The optimal scalar update $t_{ij}^{k+1}$ is determined by comparing $t_1^*$ and $t_2^*$, using the following conditions:
\[
\begin{aligned}
\text{If } 0 < t_2^* < t_1^* & \implies \rho a - \lambda > x \implies t_{ij}^{k+1} = t_1^*, \\
\text{If } t_2^* < t_1^* < 0 & \implies \rho a - \lambda < -x \implies t_{ij}^{k+1} = t_2^*,  \\
\text{If } t_2^* < 0 < t_1^* & \implies -x \le \rho a - \lambda \le x \\ 
& \implies  t_{ij}^{k+1} = \arg \min \{ g_1(t_1^*), g_2(t_2^*) \}. 
\end{aligned}
\]

\subsection{ReLU with KL divergence } \label{app:relu_kl}

The scalar problem is
$\min_{t\in\mathbb{R}}
\; \mathrm{KL}\big(x,\max(0,t)\big) + \lambda t + \tfrac{\rho}{2}(t-a)^2$.

\paragraph*{Case \(x>0\)} Because \(\mathrm{KL}(x,\cdot)\) is finite only for positive arguments when \(x>0\), the minimizer must satisfy \(t>0\) and the objective reduces to
$t - x\log t + \lambda t + \tfrac{\rho}{2}(t-a)^2$. 
Differentiating and setting to zero yields the quadratic in \(t\)
\[
\rho t^2 + (\lambda - \rho a + 1)\,t - x = 0.
\]
The unique positive root (choose the root with ‘‘\(+\)’’ in the quadratic formula) is
\begin{equation*}%\label{eq:kl_relu_pos_root}
t^* \;=\; \frac{ -(\lambda - \rho a + 1) + \sqrt{(\lambda - \rho a + 1)^2 + 4\rho x} }{2\rho}. 
\end{equation*}

\paragraph*{Case \(x=0\)} The scalar objective is %(see the definition of KL-divergence)
\[
g(t)=\max(0,t) + \lambda t + \tfrac{\rho}{2}(t-a)^2 + c.
\]
 When $x = 0$, $t$ has no restriction on its sign, which splits into two quadratics:  %for \(t>0\) and \(t\le 0\):
\[
    g(t) = \left\{ 
\begin{array}{cc}
    t + \lambda t + \tfrac{\rho}{2}(t-a)^2 & \text{ for } t>0,\\   \lambda t + \tfrac{\rho}{2}(t-a)^2 & \text{ for } t\le 0. 
\end{array}
\right. 
\]
Their unconstrained minimizers are
\[
t_1^*=\frac{\rho a - \lambda - 1}{\rho}, \qquad
t_2^*=\frac{\rho a - \lambda}{\rho},
\]
with \(t_1^*<t_2^*\), where \(s=\rho a - \lambda\). The update is determined by \(s\) as follows:
\[
t_{ij}^{k+1} =
\left\{ 
\begin{array}{cl}
t_1^* & \text{if } s > 1 \; (0 < t_1^* < t_2^*),\\ 
0     & \text{if } 0 < s \le 1 \; (t_1^* < 0 <  t_2^*), \\ 
t_2^* & \text{if } s \le 0 \; (t_1^* < t_2^* < 0). 
\end{array}
\right. 
\]

\subsection{CSF with KL divergence} \label{app:csf_kl} 

The scalar subproblem is
\begin{equation*}%\label{eq:T-update-csf-kl-scalar}
t^{k+1}_{ij}\;\in\;\argmin_{t\in\mathbb{R}}
\; g(t)\;=\; \mathrm{KL}\big(x,\,t^2\big) + \lambda t + \tfrac{\rho}{2}(t-a)^2.
\end{equation*}

\paragraph*{Case \(x>0\)}  
For \(x>0\) the KL term is always well-defined as \(t^2 \geq 0\), we just have to make sure that $t \neq 0$ and the scalar objective is, omitting additive constants, 
\[
g(t)=t^2 - x\log(t^2) + \lambda t + \tfrac{\rho}{2}(t-a)^2 .
\]
Differentiation leads to the stationary condition
\[
g'(t) = 2t - \frac{2x}{t} + \lambda + \rho(t-a) = 0,
\]
which, after rearrangement, gives a quadratic equation in \(t\). 
Solving the resulting algebraic equation yields two real candidate stationary points
\begin{equation*}%\label{eq:csf_kl_candidates}
t_{1,2}^* \;=\; \frac{\rho a - \lambda \pm \sqrt{(\lambda - \rho a)^2 + 8x(\rho+2)}}{2(\rho+2)}.
\end{equation*}
The global minimizer is one of these two candidates, the one with % set \(t_{ij}^{k+1}\) equal to the candidate with the 
the smaller objective function value. %Because \(g(t)\to+\infty\) as \(|t|\to\infty\), the global minimizer must be attained at one of these stationary candidates.

\paragraph*{Case \(x=0\)}  
When \(x=0\), the KL term reduces to \(t^2\) (up to constants) and the scalar objective is a strictly convex quadratic, 
$g(t)=t^2 + \lambda t + \tfrac{\rho}{2}(t-a)^2 + C$, 
whose unique minimizer is 
$t^* \;=\; \frac{\rho a - \lambda}{\rho + 2}$. 

\subsection{MinMax with KL-divergence} \label{app:minmax_kl}

With lower and upper bounds $p$ and $q$, the subproblem is 
\begin{equation*}
    \begin{aligned}
       \min_{t} g(t) :=  \textbf{KL}(x,\min(q, \max(p,t)))+ t\lambda + \frac{\rho}{2}(t-a)^2. 
    \end{aligned}
\end{equation*}
This splits into two equations depending on $x$ as follows 
\begin{equation*}
\begin{aligned}
 g(t) & \;=\; \min\!\big(q, \max(p,t)\big) 
        - x \log\!\big(\min(q, \max(p,t))\big)  \\
& \quad \quad +\, t\lambda 
        + \frac{\rho}{2}(t-a)^2 
        \quad \text{ for } x>0, \\ 
 g(t) & \;=\; \min\!\big(q, \max(p,t)\big) 
        + t\lambda 
        + \frac{\rho}{2}(t-a)^2 
        \quad  \text{ for } x=0.
\end{aligned}
\end{equation*}

\textbf{Case 1: $x>0$.}  When $x>0$, $t>0$, and hence 
\begin{equation*}
g(t) =
\begin{cases}
g_1(t) = q - x \log(q) + t\lambda + \dfrac{\rho}{2}(t-a)^2, 
& t > q, \\
g_2(t) = t - x \log(t) + t\lambda + \dfrac{\rho}{2}(t-a)^2, 
& p \leq t \leq q, \\
g_3(t) = p - x \log(p) + t\lambda + \dfrac{\rho}{2}(t-a)^2, 
& t < p.
\end{cases}
\end{equation*}
The three  minimizers are  
\[
\begin{aligned}
t_1^* &= \frac{\rho a - \lambda}{\rho}, && t > q, \\
t_2^* &= \frac{\rho a - \lambda - 1 + D}{2 \rho}, && p \le t \le q, \\
t_3^* &= \frac{\rho a - \lambda}{\rho}, && t < p,
\end{aligned}
\] 
where $D = \sqrt{(\lambda - \rho a + 1)^2 + 4 \rho x}$. 
The update is determined by the thresholds of $s=\rho a - \lambda$: 
\begin{itemize}
    \item When $t_1^* = t_3^* > q$ and $t_2^* > q$, $s > q\rho$ and $s > 2\rho q +1 - D$, so  
    $t^{k+1} = \argmin_{t\in \{t_1^*,q\}}  g_1(t)$. 

     \item When $t_1^* = t_3^* > q$ and $t_2^* \in  (p,q)$, $s > q\rho$ and $2\rho p +1 - D < s < 2\rho q +1 - D$, so $t^{k+1} 
     \in \{t_1^*,t_2^*,q,p\}$ depending on which has the smallest value for $g_1(t_1^*)$,  $g_1(q)$, $g_2(t_2^*)$ and $g_2(p)$. 

     \item When $t_1^* = t_3^* \in (p,q)$ and $t_2^* \in  (p,q)$, $2\rho p +1 - D < s < 2\rho q +1 - D$, so $t^{k+1} \in \{t_2^*,q,p\}$ depending on which has the smallest value for 
       $g_1(q)$, $g_2(t_2^*)$ and $g_2(p)$.
 
     \item When $t_1^* = t_3^* \in (p,q)$ and $t_2^* < p$, $\rho p< s < \rho q$ and $ s < 2\rho p +1 - D$, so $t^{k+1} = p$. 
  
     \item When $t_1^* = t_3^* < p$ and $t_2^* \in (p,q)$, $ 2\rho p +1 - D < s < 2\rho q +1 - D$, so $t^{k+1} \in \argmin_{t\in \{t_3^*,p\}}  g_3(t)$.

     \item When $t_1^* = t_3^* > q$ and $t_2^* \in  (p,q)$, $s > q\rho$ and $2\rho q +1 - D < s < 2\rho q +1 - D$, so $t^{k+1}  \in \{t_2^*,t_3^*,q,p\}$
     depending on which has the smallest value for 
      $g_3(t_3^*)$,  $g_2(q)$, $g_2(t_2^*)$ and $g_2(p)$.

     \item When $t_1^* = t_3^* \in (p,q)$ and $t_2^* > q$, $\rho p< s < \rho q$ and $ s > 2\rho q +1 - D$, so $t^{k+1} = q$. 
\end{itemize}

\textbf{Case 2: $x=0$.} The scalar objective is 
\[
g(t)= \min(q, \max(p,t)) + \lambda t + \tfrac{\rho}{2}(t-a)^2 + c,
\]
and $t$ has no restriction on its sign, which splits into 3 quadratics: 
\begin{equation*}
g(t) \;=\;
\begin{cases}
g_1(t) = q  + t\lambda + \dfrac{\rho}{2}(t-a)^2 
& \text{ for } t > q, \\
g_2(t) = t  + t\lambda + \dfrac{\rho}{2}(t-a)^2 
& \text{ for } p \leq t \leq q, \\ 
g_3(t) = p  + t\lambda + \dfrac{\rho}{2}(t-a)^2 
& \text{ for } t < p.
\end{cases}
\end{equation*} 
The unconstrained minimizers of these quadratic pieces are obtained by equating derivatives to zero. 
Thus the candidate minimizers are
\[
\; t_1^* \;=\; \frac{\rho a - \lambda}{\rho} \;,\quad
   t_2^* \;=\; \frac{\rho a - \lambda - 1}{\rho} \;,\quad
   t_3^* \;=\; \frac{\rho a - \lambda}{\rho},
\]
where $t_1^* = t_3^* > t_2^*$. 
The update is determined by the thresholds $s=\rho a - \lambda$: 
\begin{itemize}
    \item When $t_2^* < t_1^* = t_3^* < p $, then $s <\rho p$, so $t^{k+1} = t_1^*$. 
     
     \item When $t_1^* = t_3^* \in (p,q)$ and $t_2^* < p$, $s < \rho p +1$ and $\rho p < s < \rho q$, so $t^{k+1} = p$. 

     \item When $p < t_2^* < t_1^* = t_3^* < q$, 
     $\rho p < s < \rho q$ and $\rho p +1 < s < \rho q +1$, so $t^{k+1} = t_2^*$.

    \item When $t_1^* = t_3^* > q$ and $t_2^* \in  (p,q)$, $\rho p +1 < s < \rho q +1$ and $s > \rho q$, 
    so $t^{k+1} =  t_2^*$ if $g_2(t_2^*) \leq g_3(t_3^*)$, 
    and $t^{k+1} =  t_3^*$ otherwise. 

    \item When $q < t_2^* < t_1^* = t_3^* $, $ s > \rho q +1$, so  $t^{k+1} = t_3^* $.

\end{itemize}

\subsection{Modulus with KL-divergence} \label{app:modulus_kl}

The scalar subproblem is
\begin{equation*}%\label{eq:T-update-mod-kl-scalar}
t^{k+1}_{ij}\;\in\;\argmin_{t\in\mathbb{R}}
\; \mathrm{KL}\big(x,\,|t|\big) + \lambda t + \tfrac{\rho}{2}(t-a)^2 .
\end{equation*}

\paragraph*{Case \(x>0\)} The KL term enforces \(t \neq 0\). 

\noindent For the case $t>0$, the objective reduces to
$g(t)= t - x\log t + \lambda t + \tfrac{\rho}{2}(t-a)^2$, 
leading to the quadratic
\[
\rho t^2 + (\lambda - \rho a + 1)\,t - x = 0.
\]
The unique positive root is
\begin{equation}\label{eq:mod-kl-xpos}
t^*_1 \;=\; \frac{ -(\lambda - \rho a + 1) + \sqrt{(\lambda - \rho a + 1)^2 + 4\rho x} }{2\rho}. 
\end{equation} 
For the case when $t<0$, the objective reduces to
\[
g(t)= -t - x\log (-t) + \lambda t + \tfrac{\rho}{2}(t-a)^2, \qquad t<0,
\]
leading to the quadratic
$\rho t^2 + (\lambda - \rho a - 1)\,t - x = 0$.
The unique negative root is
\begin{equation}\label{eq:mod-kl-xneg}
t^*_2 \;=\; \frac{ -(\lambda - \rho a - 1) - \sqrt{(\lambda - \rho a - 1)^2 + 4\rho x} }{2\rho}. 
\end{equation}
The minimizer is chosen after comparing the objective values $g(t_1^*)$ and $g(t_2^*)$.
\paragraph*{Case \(x=0\)} The scalar objective is 
$g(t)= |t| + \lambda t + \tfrac{\rho}{2}(t-a)^2$, 
which splits into two convex quadratics:
\[
g(t) = \left\{ 
\begin{array}{cl}
 t + \lambda t + \tfrac{\rho}{2}(t-a)^2    & \text{ for }  t>0,  \\
  -t + \lambda t + \tfrac{\rho}{2}(t-a)^2    & \text{ for } t\leq 0. 
\end{array}
\right. 
\]
Their unconstrained minimizers are
\[
t_1^*=\frac{s-1}{\rho}, \qquad t_2^*=\frac{s+1}{\rho},\qquad \text{ with } s=\rho a - \lambda.
\]
The optimal update is determined by the threshold on \(s\):
\[
t_{ij}^{k+1} =
\begin{cases}
t_1^*, & \text{if } s>1 \quad (0<t_1^*<t_2^*),\\[4pt]
0,     & \text{if } -1 \le s \le 1 \quad (t_1^* \le 0 \le t_2^*),\\[4pt]
t_2^*, & \text{if } s<-1 \quad (t_1^*<t_2^*<0).
\end{cases}
\]

\subsection{ReLU with $\ell_1$ norm} \label{app:relu_l1}

The update rule is obtained by solving the following scalar optimization problem:  
\begin{equation*} %\label{eq:t-update-l1-matrix }
    \min_{t} \; |x - \max(0,t)| + t\lambda + \frac{\rho}{2}(t-a)^2 . 
\end{equation*} 
This can be split into two regions considering $x\geq 0$ because the data matrix in the ReLU case is always nonnegative: 
\[
\begin{split}
g_1(t) &= x + t\lambda + \frac{\rho}{2}(t-a)^2, \quad t \leq 0, \\
g_2(t) &= |x-t| + t\lambda + \frac{\rho}{2}(t-a)^2, \quad t > 0 .
\end{split}
\] 
The minimizers of these two cases are denoted by $t_1^*$ and $t_2^*$ respectively.   
For $t \leq 0$, the function $g_1(t)$ is quadratic and attains its minimum at
$t_1^* = \frac{\rho a - \lambda}{\rho}$.
For $t > 0$, the minimizer can be expressed in compact form as
\[
t_2^* = \max\!\Bigg( \min\!\left( \frac{\rho a - \lambda + 1}{\rho}, \, x \right), \, \frac{\rho a - \lambda - 1}{\rho} \Bigg).
\]

Depending on the value of $\rho a - \lambda$, the selection is as follows: 
\begin{itemize}
    \item If $\rho a - \lambda > 1$, both $t_1^*$ and $t_2^*$ are positive, so $t^* = t_2^*$.

    \item If $\rho a - \lambda < -1$, both $t_1^*$ and $t_2^*$ are negative, so $t^* = t_1^*$.

    \item If $0 < \rho a - \lambda < 1$, $t_1^* < 0 < t_2^*$, so $t^* = 0$.

    \item If $-1 < \rho a - \lambda < 0$, $t_2^* > 0 > t_1^*$, so $t^* \in \{t_1^*, t_2^*\}$, whichever has a lower objective function value.
\end{itemize}

\subsection{CSF with $\ell_1$ norm} \label{app:csf_l1}

 The scalar problem is
\[
\min_{t\in\mathbb{R}} \; |x - t^2| + \lambda t + \frac{\rho}{2}(t-a)^2.
\]
This splits by region into two quadratics (plus a kink at $t=\pm\sqrt{x}$):
\[
\begin{aligned}
g_1(t) &= x - t^2 + \lambda t + \tfrac{\rho}{2}(t-a)^2, \quad &\text{if } t^2\le x,\\
g_2(t) &= t^2 - x + \lambda t + \tfrac{\rho}{2}(t-a)^2, \quad &\text{if } t^2> x.
\end{aligned}
\]
Their unconstrained minimizers are:
\[
t_1^*=\frac{s}{\rho-2}\quad(\text{for }g_1),\qquad
t_2^*=\frac{s}{\rho+2}\quad(\text{for }g_2),
\]
where s = $\rho a - \lambda$,  together with the boundary candidates \(t_b^\pm=\pm\sqrt{x}\) at the kink. 
To ensure $g_1$ is convex, we recommend \(\rho\ge 2\); when \(\rho<2\), the minimum of $g_1$ over $\{t:\,t^2\le x\}$ occurs on the boundary $t=\pm\sqrt{x}$.

When $\rho\ge 2$, feasibility of the stationary points is characterized by simple thresholds in $|s|$: 
\[
\text{$t_1^*$ feasible} \;\Longleftrightarrow\; (t_1^*)^2\le x 
\;\Longleftrightarrow\; |s|\le (\rho-2)\sqrt{x},
\]
\[
\text{$t_2^*$ feasible} \;\Longleftrightarrow\; (t_2^*)^2\ge x 
\;\Longleftrightarrow\; |s|\ge (\rho+2)\sqrt{x}.
\]
Hence, 
\begin{itemize}
  \item \textbf{Case A:} $|s|\ge (\rho+2)\sqrt{x}$ \;(\(t_2^*\) feasible in $t^2>x$). 
  Evaluate $g_2(t_2^*)$ and $g(t_b^\pm)$, and set $t^{k+1}$ to the argument with the smaller value.
  
  \item \textbf{Case B:} $|s|\le (\rho-2)\sqrt{x}$ \;(\(t_1^*\) feasible in $t^2\le x$). 
  Evaluate $g_1(t_1^*)$ and $g(t_b^\pm)$, and set $t^{k+1}$ to the smaller. 
  
  \item \textbf{Case C:} $(\rho-2)\sqrt{x}<|s|<(\rho+2)\sqrt{x}$ \;(no feasible stationary point). 
  The minimizer lies at the kink: set 
  \(t^{k+1}=\argmin_{t \in \{t_b^-, t_b^+\}} g(t)\). 
\end{itemize}

When $\rho<2$, skip $t_1^*$ and select from $\{t_2^*,\,t_b^\pm\}$ using the same evaluation rule.

\subsection{MinMax with $\ell_1$ norm} \label{app:minmax_l1}

 The scalar optimization problem is    
\begin{equation*} %\label{eq:t-update-l1-matrix }
    \min_{t} \; |x - \min(q, \max(p,t))| + t\lambda + \frac{\rho}{2}(t-a)^2 . 
\end{equation*} 
This can be split into three regions considering $x\geq 0$ because the data is assumed to lie in the interval $[p,q]$: 
\[
\begin{split}
g_1(t) &= |x-p| + t\lambda + \frac{\rho}{2}(t-a)^2, \quad t < p, \\
g_2(t) &= |x-t| + t\lambda + \frac{\rho}{2}(t-a)^2, \quad p \leq t \leq q , \\ 
g_3(t) &= |x-q| + t\lambda + \frac{\rho}{2}(t-a)^2, \quad t >q. 
\end{split}
\] 
The minimizers of these three cases are denoted by $t_1^*$, $t_2^*$ and $t_3^*$ respectively.   
For $t < p$ and $t>q$ the functions $g_1(t)$ and $g_3(t)$ are quadratic functions in $t$ and attain their  minimum at
\[
t_1^* = \frac{\rho a - \lambda}{\rho} = t_3^*.
\] 
For $ p \leq t \leq q$, the minimizer of $g_2(t)$ can be expressed in compact form as
\[
t_2^* = \max\!\Bigg( \min\!\left( \frac{\rho a - \lambda + 1}{\rho}, \, x \right), \, \frac{\rho a - \lambda - 1}{\rho} \Bigg).
\]
 The update is determined by the thresholds of $s=\rho a - \lambda$: 
\begin{itemize}

\item When $q < t_2^* , t_1^* = t_3^* $, $ s > \rho q +1$, so $t^{k+1} = t_3^*$. 

 \item When $p < t_2^* , t_1^* = t_3^* < q$, $\rho p < s < \rho q$, $\rho p +1 < s < \rho q +1$ and $\rho p -1 < s < \rho q - 1$, so $t^{k+1} = t_2^*$. 
 
    \item When $t_2^*, t_1^* = t_3^* < p$, 
    $s <\rho p -1 $, so $t^{k+1} = t_1^*$.

     \item When $t_1^* = t_3^* < p$ and $t_2^* \in  (p,q)$, then $\rho p +1 < s < \rho q +1$, $\rho p -1 < s < \rho q -1$ and $s < \rho p$, 
     so $t^{k+1} \in \{t_2^*,t_3^*\}$ depending on which value,  $g_1(t_1^*)$ or $g_2(t_2^*)$, is smaller. 

     \item When $t_1^* = t_3^* \in (p,q)$ and $t_2^* > q$, then $ \rho p < s < \rho q$, $ s > \rho q -1 $ and $ s > \rho q + 1$, so $t^{k+1} = q$.

     \item When $t_1^* = t_3^* > q$ and $t_2^* \in  (p,q)$, then $\rho p +1 < s < \rho q +1$ and $s > \rho q$, so $t^{k+1} \in \{ t_2^* , t_3^*\}$ depending on which value, $g_2(t_1^*)$ or $g_3(t_2^*)$, is smaller.

    \item When $t_1^* = t_3^* \in (p,q)$ and $t_2^* < p$, then $ \rho p < s < \rho q$, $ s < \rho p -1 $ and $ s < \rho p + 1 $, so $t^{k+1} = p$.

\end{itemize}

\subsection{Modulus with $\ell_1$ norm } \label{app:modulus_l1}

The scalar optimization problem is  
\begin{equation*} %\label{eq:t-update-l1-matrix }
    \min_{t} \; |x - |t| | + t\lambda + \frac{\rho}{2}(t-a)^2 . 
\end{equation*} 
This can be split into two regions considering $x\geq 0$ because the data matrix in the Modulus is assumed to be   nonnegative: 
\[
\begin{split}
g_1(t) &= |x - t| + t \lambda + \frac{\rho}{2}(t-a)^2, \quad t >0 , \\
g_2(t) &= |x+t| + t\lambda + \frac{\rho}{2}(t-a)^2, \quad t \leq 0 .
\end{split}
\] 
For $t > 0$, the minimizer can be expressed as
\[
t_1^* = \max\!\Bigg( \min\!\left( \frac{\rho a - \lambda + 1}{\rho}, \, x \right), \, \frac{\rho a - \lambda - 1}{\rho} \Bigg).
\] 
For $t \leq 0$, the minimizer is 
\[
t_2^* = \max\!\Bigg( \min\!\left( \frac{\rho a - \lambda + 1}{\rho}, \, -x \right), \, \frac{\rho a - \lambda - 1}{\rho} \Bigg).
\]
There are three cases: 
\begin{enumerate}
    \item \textbf{Case 1:} $t_2^* , t_1^* < 0$ (i.e., $\rho a - \lambda < -1$), so 
    $t^{k+1} = t_2^*$.

    \item \textbf{Case 2:} $0 < t_2^* , t_1^*$ (i.e., $ \rho a - \lambda > 1$), so 
    $t^{k+1} = t_1^*$.

    \item \textbf{Case 3:} $t_2^* < 0 < t_1^*$ (i.e., $-1 <  \rho a - \lambda < 1$), so 
    $ 
    t^{k+1} = t_1^*$ if $g_1(t_1^*) < g_2(t_2^*)$, and $= t_2^*$ otherwise.

\end{enumerate}

\section{Hardness of NMDs}
\label{sup:hardness}

As illustrated on Figure 2, the CSF model fails to converge to accurate solutions and is unable to recover the underlying factor matrices $W$ and $H$ used to generate the data matrix $X = f(WH)$. This behavior can be understood by examining the structure of the CSF
nonlinearity, $f(t)=t^2$. 
In our synthetic setting, the entries of $W$ and $H$ are sampled from a Gaussian
distribution, so $WH$ typically contains both positive
and negative values (roughly half are positive and half are negative).
Applying the CSF nonlinearity entrywise discards all sign information and maps all values to the nonnegative domain. Consequently, even when an exact representation exists,
recovering factors $(W,H)$ such that $X=(WH)^{.2}$ requires the ADMM algorithm to
implicitly resolve a large set of sign ambiguities,
which implicitly requires to solve a hard combinatorial problem. 
In fact, computing the smallest $r\in\mathbb{N}$ such that $X=Y^{.2}$ for some $\mathrm{rank}(Y)\le r$ amounts to computing the \emph{square-root rank} of $X$, an NP-hard problem; see, e.g.,~\cite{fawzi2015positive}. 
As a result, our ADMM algorithm is likely to get stuck in bad local minima. 
Similar observations have been reported in~\cite{lefebvre2024component}, where the difficulties associated with sign ambiguities are discussed in greater detail. A similar observation applies to the Modulus nonlinear function. 

To shed further light on the behavior observed for the CSF model, we perform additional synthetic experiments with more favorable settings in order to assess whether the previous poor performances of CSF are due to the hardness of the problem rather than a limitation of our ADMM framework. 
\begin{itemize}
    \item With $m=100$, $n=80$, and $r=5$, we generate $X\in\mathbb{R}^{m\times n}$ by drawing $W\in\mathbb{R}^{m\times r}$ and $H\in\mathbb{R}^{r\times n}$ with i.i.d.\ entries from the uniform distribution on $[0,1]$ (using the MATLAB command \texttt{rand}) and setting $X=(WH)^{.2}$.
    The sign ambiguity is now trivial since, by construction, there exist optimal nonnegative factors $W$ and $H$ (although the ADMM algorithm does not use this information). 
    On the top of Figure~\ref{fig:csf-rand}, we observe the evolution of the average objective value over 10 runs for 15 iterations. In contrast to the Gaussian case, the algorithm converges to small objective function values (around 0.01\% relative error or below), typically within the first 10 iterations.
    
    \item Let us come back to the Gaussian case, meaning the entries of $W$ and $H$ are generated using the Gaussian distribution, and $X = (WH).^2$, but consider small instances, namely
    $m=n=10$, $r=2$. % preserves the original Gaussian distribution for the entries of $W\in\mathbb{R}^{m\times r}$ and $H\in\mathbb{R}^{r\times n}$. 
    This smaller problem reduces the importance of the combinatorial complexity associated with sign patterns in the product $WH$.
    On the bottom of Figure~\ref{fig:csf-rand}, we observe the evolution of the average objective value over 10 runs for 15 iterations. The fast decrease of the objective indicates that, in small dimensions, ADMM can reach accurate solutions reliably (below 1\%).  
    This supports the interpretation that the poor behavior observed at larger scales is mainly due to the complexity of the problem, rather than by an intrinsic deficiency of the ADMM scheme.
\end{itemize}

\begin{figure}[!htbp]
\centering
\begin{minipage}[t]{\columnwidth}
        \centering
        \begin{tikzpicture}[scale=0.8]
            \begin{axis}[
                ymode=log,
                xlabel={Iterations},
                ylabel = {Average objective},
                grid=both,
                legend style={at={(0.27, 0.4)}, anchor=north, legend columns=1, font=\Large}, 
                tick label style={font=\Large},
                xlabel style={font=\Large}, 
                ylabel style={font=\Large}
            ]
                
                \addplot[line width=2pt, color=violet] table [x expr=\coordindex,y=y1, col sep=space] {CSFresults.txt};
                \addlegendentry{Frobenius}
    
                \addplot[line width=1.5pt, color=blue, mark=triangle*,
  mark options={scale=0.8, fill=OliveGreen},
  mark repeat=1,color=blue] table [x expr=\coordindex,y=y2, col sep=space] {CSFresults.txt};
                \addlegendentry{$\ell_1$}

                \addplot[line width=2pt, mark=*,
  mark options={scale=0.8, fill=OliveGreen},
  mark repeat=1,color=OliveGreen] table [x expr=\coordindex,y=y3, col sep=space] {CSFresults.txt};
                \addlegendentry{KL}

            \end{axis}
        \end{tikzpicture}
\vspace{0.5em}

\textbf{(a)} Uniform distribution in [0,1] of the entries of $W$, $H$, $m=100$, $n=80$, $r= 5$.  
\end{minipage}

\vspace{1em}

\centering  
\begin{minipage}[t]{\columnwidth}
        \centering 
        \begin{tikzpicture}[scale=0.8]
            \begin{axis}[
                ymode=log,
                xlabel={Iterations},
                ylabel = {Average objective},
                grid=both,
                legend style={at={(0.4, 0.4)}, anchor=north, legend columns=1, font=\Large}, 
                tick label style={font=\Large},
                xlabel style={font=\Large}, 
                ylabel style={font=\Large}
            ]
                
                \addplot[line width=2pt, color=violet] table [x expr=\coordindex,y=y1, col sep=space] {CSF_randn_results.txt};
                \addlegendentry{Frobenius}
    
                \addplot[line width=1.5pt, color=blue, mark=triangle*,
  mark options={scale=0.8, fill=OliveGreen},
  mark repeat=1,color=blue] table [x expr=\coordindex,y=y2, col sep=space] {CSF_randn_results.txt};
                \addlegendentry{$\ell_1$}

                \addplot[line width=2pt, mark=*,
  mark options={scale=0.8, fill=OliveGreen},
  mark repeat=1,color=OliveGreen] table [x expr=\coordindex,y=y3, col sep=space] {CSF_randn_results.txt};
                \addlegendentry{KL}
             
            \end{axis}
        \end{tikzpicture}
        
        \textbf{(b)} Gaussian distribution of the entries of $W$, $H$, $m=n=10$, $r=2$.  
\end{minipage}

\caption{Average objective value as a function of the iteration number for the CSF model under two data-generation regimes: (a) entries of $W$ and $H$ follow a uniform distribution in [0,1] and $m=100$, $n=80$ and $r=5$, 
and (b) entries of $W$ and $H$ follow a standard Gaussian distribution   $m=n=10$ and $r=2$. 
}
\label{fig:csf-rand}
\end{figure}

Interestingly, in the two simple cases above, the KL divergence allows ADMM to converge to  machine precision  errors.
It is unclear to us why this is the case and is a question for future research. 

\section{Ablation study: adaptive versus fixed penalty parameter}
\label{sup:rho}

\revision{A key component of the proposed ADMM framework is the adaptive update of the penalty parameter $\rho$, which balances the primal and dual residuals during optimization (see Subsection IV-C). In this section, we evaluate the impact of this adaptive strategy by comparing it against fixed values of $\rho \in \{0.01, 0.1, 1, 10, 100\}$ across the ReLU and MinMax models with all three loss functions.

We use the same synthetic setup as in Section V: $m = 100$, $n = 80$, $r = 5$, with $10^3$ iterations and results averaged over 10 independent runs. All configurations share the same initialization for each run to ensure a fair comparison. The results are shown in Figure~\ref{fig:rho_ablation}.

\begin{figure}[!htbp]
\centering
 
\begin{minipage}[t]{0.48\columnwidth}
    \centering
    \begin{tikzpicture}[scale=0.72]
        \begin{axis}[
            xlabel={Iterations},
            ylabel={Average objective},
            grid=both,
            legend style={at={(0.8,0.98)}, anchor=north east, legend columns=1, font=\large},
            tick label style={font=\Large},
            xlabel style={font=\Large},
            ylabel style={font=\Large},
            title={\Large ReLU + Frobenius},
            title style={at={(0.5,1.02)}},
        ]
            \addplot[line width=1.5pt, dashed, color={rgb,255:red,204;green,0;blue,204}, mark=v, mark repeat=100] table [x expr=\coordindex, y=y1, col sep=space] {rho_ablation_ReLU_Frobenius.txt};
            \addlegendentry{$\rho{=}0.01$}
            \addplot[line width=1.5pt, dashed, color=cyan, mark=diamond, mark repeat=100] table [x expr=\coordindex, y=y2, col sep=space] {rho_ablation_ReLU_Frobenius.txt};
            \addlegendentry{$\rho{=}0.1$}
            \addplot[line width=1.5pt, dashed, color=gray, mark=pentagon, mark repeat=100] table [x expr=\coordindex, y=y3, col sep=space] {rho_ablation_ReLU_Frobenius.txt};
            \addlegendentry{$\rho{=}1$}
            \addplot[line width=1.5pt, dashed, color=orange, mark=otimes, mark repeat=100] table [x expr=\coordindex, y=y4, col sep=space] {rho_ablation_ReLU_Frobenius.txt};
            \addlegendentry{$\rho{=}10$}
            \addplot[line width=1.5pt, dashed, color={rgb,255:red,153;green,51;blue,204}, mark=triangle, mark repeat=100] table [x expr=\coordindex, y=y5, col sep=space] {rho_ablation_ReLU_Frobenius.txt};
            \addlegendentry{$\rho{=}100$}
            \addplot[line width=2pt, solid, color=blue, mark=triangle*, mark repeat=100] table [x expr=\coordindex, y=y6, col sep=space] {rho_ablation_ReLU_Frobenius.txt};
            \addlegendentry{Adaptive}
        \end{axis}
    \end{tikzpicture}
\end{minipage}
\hfill
\begin{minipage}[t]{0.48\columnwidth}
    \centering
    \begin{tikzpicture}[scale=0.72]
        \begin{axis}[
            xlabel={Iterations},
            grid=both,
            tick label style={font=\Large},
            xlabel style={font=\Large},
            ylabel style={font=\Large},
            title={\Large MinMax + Frobenius},
            title style={at={(0.5,1.02)}},
        ]
            \addplot[line width=1.5pt, dashed, color={rgb,255:red,204;green,0;blue,204}, mark=v, mark repeat=100] table [x expr=\coordindex, y=y1, col sep=space] {rho_ablation_MinMax_Frobenius.txt};
            \addplot[line width=1.5pt, dashed, color=cyan, mark=diamond, mark repeat=100] table [x expr=\coordindex, y=y2, col sep=space] {rho_ablation_MinMax_Frobenius.txt};
            \addplot[line width=1.5pt, dashed, color=gray, mark=pentagon, mark repeat=100] table [x expr=\coordindex, y=y3, col sep=space] {rho_ablation_MinMax_Frobenius.txt};
            \addplot[line width=1.5pt, dashed, color=orange, mark=otimes, mark repeat=100] table [x expr=\coordindex, y=y4, col sep=space] {rho_ablation_MinMax_Frobenius.txt};
            \addplot[line width=1.5pt, dashed, color={rgb,255:red,153;green,51;blue,204}, mark=triangle, mark repeat=100] table [x expr=\coordindex, y=y5, col sep=space] {rho_ablation_MinMax_Frobenius.txt};
            \addplot[line width=2pt, solid, color=blue, mark=triangle*, mark repeat=100] table [x expr=\coordindex, y=y6, col sep=space] {rho_ablation_MinMax_Frobenius.txt};
        \end{axis}
    \end{tikzpicture}
\end{minipage}
 
\vspace{0.8em}
 
\begin{minipage}[t]{0.48\columnwidth}
    \centering
    \begin{tikzpicture}[scale=0.72]
        \begin{axis}[
            xlabel={Iterations},
            ylabel={Average objective},
            grid=both,
            tick label style={font=\Large},
            xlabel style={font=\Large},
            ylabel style={font=\Large},
            title={\Large ReLU + $\ell_1$},
            title style={at={(0.5,1.02)}},
        ]
            \addplot[line width=1.5pt, dashed, color={rgb,255:red,204;green,0;blue,204}, mark=v, mark repeat=100] table [x expr=\coordindex, y=y1, col sep=space] {rho_ablation_ReLU_L1.txt};
            \addplot[line width=1.5pt, dashed, color=cyan, mark=diamond, mark repeat=100] table [x expr=\coordindex, y=y2, col sep=space] {rho_ablation_ReLU_L1.txt};
            \addplot[line width=1.5pt, dashed, color=gray, mark=pentagon, mark repeat=100] table [x expr=\coordindex, y=y3, col sep=space] {rho_ablation_ReLU_L1.txt};
            \addplot[line width=1.5pt, dashed, color=orange, mark=otimes, mark repeat=100] table [x expr=\coordindex, y=y4, col sep=space] {rho_ablation_ReLU_L1.txt};
            \addplot[line width=1.5pt, dashed, color={rgb,255:red,153;green,51;blue,204}, mark=triangle, mark repeat=100] table [x expr=\coordindex, y=y5, col sep=space] {rho_ablation_ReLU_L1.txt};
            \addplot[line width=2pt, solid, color=blue, mark=triangle*, mark repeat=100] table [x expr=\coordindex, y=y6, col sep=space] {rho_ablation_ReLU_L1.txt};
        \end{axis}
    \end{tikzpicture}
\end{minipage}
\hfill
\begin{minipage}[t]{0.48\columnwidth}
    \centering
    \begin{tikzpicture}[scale=0.72]
        \begin{axis}[
            xlabel={Iterations},
            grid=both,
            tick label style={font=\Large},
            xlabel style={font=\Large},
            ylabel style={font=\Large},
            title={\Large MinMax + $\ell_1$},
            title style={at={(0.5,1.02)}},
        ]
            \addplot[line width=1.5pt, dashed, color={rgb,255:red,204;green,0;blue,204}, mark=v, mark repeat=100] table [x expr=\coordindex, y=y1, col sep=space] {rho_ablation_MinMax_L1.txt};
            \addplot[line width=1.5pt, dashed, color=cyan, mark=diamond, mark repeat=100] table [x expr=\coordindex, y=y2, col sep=space] {rho_ablation_MinMax_L1.txt};
            \addplot[line width=1.5pt, dashed, color=gray, mark=pentagon, mark repeat=100] table [x expr=\coordindex, y=y3, col sep=space] {rho_ablation_MinMax_L1.txt};
            \addplot[line width=1.5pt, dashed, color=orange, mark=otimes, mark repeat=100] table [x expr=\coordindex, y=y4, col sep=space] {rho_ablation_MinMax_L1.txt};
            \addplot[line width=1.5pt, dashed, color={rgb,255:red,153;green,51;blue,204}, mark=triangle, mark repeat=100] table [x expr=\coordindex, y=y5, col sep=space] {rho_ablation_MinMax_L1.txt};
            \addplot[line width=2pt, solid, color=blue, mark=triangle*, mark repeat=100] table [x expr=\coordindex, y=y6, col sep=space] {rho_ablation_MinMax_L1.txt};
        \end{axis}
    \end{tikzpicture}
\end{minipage}
 
\vspace{0.8em}
 
\begin{minipage}[t]{0.48\columnwidth}
    \centering
    \begin{tikzpicture}[scale=0.72]
        \begin{axis}[
            xlabel={Iterations},
            ylabel={Average objective},
            grid=both,
            tick label style={font=\Large},
            xlabel style={font=\Large},
            ylabel style={font=\Large},
            title={\Large ReLU + KL},
            title style={at={(0.5,1.02)}},
        ]
            \addplot[line width=1.5pt, dashed, color={rgb,255:red,204;green,0;blue,204}, mark=v, mark repeat=100] table [x expr=\coordindex, y=y1, col sep=space] {rho_ablation_ReLU_KL.txt};
            \addplot[line width=1.5pt, dashed, color=cyan, mark=diamond, mark repeat=100] table [x expr=\coordindex, y=y2, col sep=space] {rho_ablation_ReLU_KL.txt};
            \addplot[line width=1.5pt, dashed, color=gray, mark=pentagon, mark repeat=100] table [x expr=\coordindex, y=y3, col sep=space] {rho_ablation_ReLU_KL.txt};
            \addplot[line width=1.5pt, dashed, color=orange, mark=otimes, mark repeat=100] table [x expr=\coordindex, y=y4, col sep=space] {rho_ablation_ReLU_KL.txt};
            \addplot[line width=1.5pt, dashed, color={rgb,255:red,153;green,51;blue,204}, mark=triangle, mark repeat=100] table [x expr=\coordindex, y=y5, col sep=space] {rho_ablation_ReLU_KL.txt};
            \addplot[line width=2pt, solid, color=blue, mark=triangle*, mark repeat=100] table [x expr=\coordindex, y=y6, col sep=space] {rho_ablation_ReLU_KL.txt};
        \end{axis}
    \end{tikzpicture}
\end{minipage}
\hfill
\begin{minipage}[t]{0.48\columnwidth}
    \centering
    \begin{tikzpicture}[scale=0.72]
        \begin{axis}[
            xlabel={Iterations},
            grid=both,
            tick label style={font=\Large},
            xlabel style={font=\Large},
            ylabel style={font=\Large},
            title={\Large MinMax + KL},
            title style={at={(0.5,1.02)}},
        ]
            \addplot[line width=1.5pt, dashed, color={rgb,255:red,204;green,0;blue,204}, mark=v, mark repeat=100] table [x expr=\coordindex, y=y1, col sep=space] {rho_ablation_MinMax_KL.txt};
            \addplot[line width=1.5pt, dashed, color=cyan, mark=diamond, mark repeat=100] table [x expr=\coordindex, y=y2, col sep=space] {rho_ablation_MinMax_KL.txt};
            \addplot[line width=1.5pt, dashed, color=gray, mark=pentagon, mark repeat=100] table [x expr=\coordindex, y=y3, col sep=space] {rho_ablation_MinMax_KL.txt};
            \addplot[line width=1.5pt, dashed, color=orange, mark=otimes, mark repeat=100] table [x expr=\coordindex, y=y4, col sep=space] {rho_ablation_MinMax_KL.txt};
            \addplot[line width=1.5pt, dashed, color={rgb,255:red,153;green,51;blue,204}, mark=triangle, mark repeat=100] table [x expr=\coordindex, y=y5, col sep=space] {rho_ablation_MinMax_KL.txt};
            \addplot[line width=2pt, solid, color=blue, mark=triangle*, mark repeat=100] table [x expr=\coordindex, y=y6, col sep=space] {rho_ablation_MinMax_KL.txt};
        \end{axis}
    \end{tikzpicture}
\end{minipage}
 
\caption{\revision{Ablation study comparing the adaptive penalty parameter strategy (solid blue) against fixed $\rho$ values (dashed) on synthetic data ($m{=}100$, $n{=}80$, $r{=}5$, averaged over 10 runs). Left column: ReLU model. Right column: MinMax model. Rows correspond to the Frobenius (top), $\ell_1$ (middle), and KL divergence (bottom) loss functions. The legend in the top-left subplot applies to all panels.}}
\label{fig:rho_ablation}
\end{figure}

Several observations emerge from Figure~\ref{fig:rho_ablation}. 
First, large fixed values of $\rho$ ($\rho = 10$ and $\rho = 100$) 
consistently lead to slow convergence across all six combinations, 
as over-penalizing the augmented Lagrangian constraint slows progress 
on the original objective. Second, no single small fixed value of 
$\rho$ performs well universally. For instance, $\rho = 0.01$ 
converges quickly for ReLU but stagnates at a high objective value 
for MinMax + KL divergence. The adaptive strategy (solid blue) avoids 
these extremes by balancing the primal and dual residuals 
automatically. While it is not always the absolute fastest, for 
example, $\rho = 0.01$ converges slightly faster on MinMax + 
Frobenius, it consistently ranks among the top-performing 
configurations across all combinations and never suffers the 
catastrophic stagnation observed with poorly chosen fixed values. 
This robustness eliminates the need to tune $\rho$ for each specific 
model/loss combination. }

\revisionSP{

\section{Convergence Analysis}
\label{sec:convergence}

In this section, we establish subsequential convergence of ADMM to stationary points for the proposed ADMM
framework under some specific assumptions; see Theorem~\ref{thm:convergence}. 
As discussed in Section~4, the coupling
constraint $T = WH$ is bilinear rather than linear, so existing
convergence theory for nonconvex
ADMM~\cite{hong2016convergence,wang2019global} does not apply directly.
The key observation enabling our analysis is that, although the
constraint is bilinear in $(W,H)$, the block $T$ enters it with the
identity coefficient, since $T=WH$. 
In ADMM, it is precisely the coefficient
of the last-updated block that must be surjective, since this is what
allows the dual update to be controlled by a primal quantity. The
bilinearity affects only the $W$- and $H$-blocks, and is handled through 
strong convexity of the corresponding subproblems.

\paragraph*{Assumptions and ADMM updates} 

In order to prove convergence, we will have to make two  important assumptions: 
\begin{itemize}
    \item The function $d(X,f(T))$ in variable $T$ is continuously differentiable and Lipschitz smooth (L-smooth), that is, its gradient is Lipschitz continuous. This assumption is not directly satisfied by our models: 
    $\max(0,\cdot)$, $\min(b,\max(a,\cdot)$, $|\cdot|$, and $\|\cdot\|_1$ are not differentiable, the KL divergence is not L-smooth at 0, and $f(t) = \| x - t^2 \|_2^2$ is not globally L-smooth. 
    However, \begin{itemize}
        \item It would be relatively straightforward to use existing smoothings of these objectives; see, e.g., \cite{chen1996class} and many follow-up works. For example, one can replace the ReLU by the softplus function, $f(t) = \frac{1}{\beta}   \log(1+e^{\beta t})$, where increasing $\beta > 0$ improves the approximation. 
        Together with the Frobenius norm, it would satisfy the above assumption. Similarly, MinMax can be approximated by $f_\beta(t) = a
+ \frac{1}{\beta}\log\left(1+e^{\beta(t-a)}\right)
- \frac{1}{\beta}\log\left(1+e^{\beta(t-b)}\right)$. 
Another example is to replace the Frobenius norm by $\sqrt{\|X-T\|_F^2 + \epsilon^2} - \epsilon$ which combined with the nonlinearity $f(t) = t^2$ would satisfy the assumption above.  

        \item Our objectives satisfy these assumptions almost everywhere and/or locally.

\item Some existing NMDs satisfy these assumptions; in particular with the sigmoid nonlinearity $f(t) = \frac{1}{1+e^{-t}}$ combined with the Frobenius norm. This NMD model was successfully used in~\cite{awari2024coordinate} to approximate and complete bounded data. 
        
\end{itemize}

    \item The penalty parameter $\rho$ is frozen after a finite number of iterations. In fact, our adaptive strategy, although very efficient in practice (see Section~\ref{sup:rho}), would be significantly harder to analyze theoretically. 
    
\end{itemize}
Extending our proof without these assumptions is an active area of research in the ADMM literature, and out of the scope of this paper.  

Let $\ell(T) := d\big(X, f(T)\big)$ and note that, by our  separability
assumption,
$\ell(T) = \sum_{i,j} \ell_{ij}(T_{ij})$. Let $A := WH$ and define the
regularized augmented Lagrangian 
\begin{equation*}
\widetilde{L}_\rho(W,H,T,\Lambda)
:= \ell(T) + \langle \Lambda, T - WH\rangle 
+ \frac{\rho}{2}\|T - WH\|_F^2
+ \frac{\mu}{2}\big(\|W\|_F^2 + \|H\|_F^2\big),
\end{equation*}
where $\mu > 0$ is a fixed regularization parameter. It is equivalent to the use $\varepsilon_H = \varepsilon_W = \mu$ in our implementation. 
The iterations are
\begin{equation}
\label{eq:mod_updates}
\begin{aligned}
W^{k+1} &= \argmin_W \ \widetilde{L}_\rho(W, H^k, T^k, \Lambda^k), \\
H^{k+1} &= \argmin_H \ \widetilde{L}_\rho(W^{k+1}, H, T^k, \Lambda^k), \\
T^{k+1} &= \argmin_T \ \widetilde{L}_\rho(W^{k+1}, H^{k+1}, T, \Lambda^k), \\
\Lambda^{k+1} &= \Lambda^k + \rho\big(T^{k+1} - W^{k+1}H^{k+1}\big). 
\end{aligned}
\end{equation}
All updates have closed-form solutions, as explained in our paper.  
Let us formalize our assumptions. 
\begin{assumption}
\label{ass:main}
The following hold.
\begin{enumerate}[label=(A\arabic*), leftmargin=*]
\item \label{ass:nonneg} $\ell(T) \ge 0$ for all
$T \in \mathbb{R}^{m\times n}$.
\item \label{ass:smooth} $\ell$ is continuously differentiable and
$\nabla \ell$ is $L$-Lipschitz continuous on
$\mathbb{R}^{m\times n}$.
\item \label{ass:rho} The penalty parameter is fixed, with
$\rho > 2L$, and $\mu > 0$.
\end{enumerate}
\end{assumption}

Assumption~\ref{ass:nonneg} holds for all 12 loss functions
considered in this paper. Assumption~\ref{ass:rho} can be  enforced in
practice by running the adaptive rule for a fixed
number $K_0$ of burn-in iterations and freezing $\rho$ thereafter. 
Assumption~\ref{ass:smooth} is the substantive restriction discussed above. 

Let us denote $A^k := W^k H^k$ and $R^k := T^k - A^k$ for the primal residual, we collect the iterates into
$\mathcal{X}^k := (W^k,H^k,T^k,\Lambda^k)$, and we abbreviate
$\widetilde{L}^k := \widetilde{L}_\rho(\mathcal{X}^k)$. Increments are
written $\Delta W^k := W^{k+1}-W^k$, and analogously for
$\Delta H^k$, $\Delta T^k$ and $\Delta \Lambda^k$.

In the following, we follow a standard strategy to prove subsequential convergence: sufficient decrease of $\tilde L$, boundedness of $\tilde L$ and of the iterates $\mathcal{X}^k$, and finally subsequential convergence to stationary points.

\paragraph*{Sufficient decrease} In order to prove a sufficient decrease lemma (Lemma~\ref{lem:decrease}), we need the following lemma. 

\begin{lemma}[Dual step controlled by primal step]
\label{lem:dual}
Under Assumption~\ref{ass:main}, for every $k \ge 1$,
\begin{equation*}
\Lambda^{k} = -\nabla \ell(T^{k}),
\quad
\|\Delta\Lambda^k\|_F \le L\,\|\Delta T^k\|_F .
\end{equation*}
\end{lemma}

\begin{proof}
The first-order optimality condition of the $T$-subproblem reads
\begin{equation}
\label{eq:Topt}
\nabla\ell(T^{k+1}) + \Lambda^k + \rho R^{k+1} = 0 ,
\end{equation}
recalling $R^{k+1} = T^{k+1} - A^{k+1}$.
By the dual update,
$\Lambda^{k+1} = \Lambda^k + \rho(T^{k+1}-A^{k+1})$, so \eqref{eq:Topt}
becomes $\nabla\ell(T^{k+1}) + \Lambda^{k+1} = 0$, which proves
$\Lambda^{k} = -\nabla\ell(T^{k})$ for every $k \ge 1$. The second
claim follows from \ref{ass:smooth} for every $k \ge 1$.
\end{proof}

\begin{lemma}[Sufficient decrease]
\label{lem:decrease}
Under Assumption~\ref{ass:main}, for every $k \ge 1$,
\begin{equation}
\label{eq:decrease}
\widetilde{L}^{k} - \widetilde{L}^{k+1} \;\ge\;
\tfrac{\mu}{2}\|\Delta W^k\|_F^2
+ \tfrac{\mu}{2}\|\Delta H^k\|_F^2 
+ \delta\,\|\Delta T^k\|_F^2,
\end{equation}
where $\delta := \tfrac{\rho - L}{2} - \tfrac{L^2}{\rho} > 0$.
\end{lemma}

\begin{proof}
We decompose the decrease across the four updates. Let us denote 
\begin{equation*}
\begin{aligned}
\mathcal{L}_1^k &:= \widetilde{L}_\rho(W^k,H^k,T^k,\Lambda^k), \\
\mathcal{L}_2^k &:= \widetilde{L}_\rho(W^{k+1},H^k,T^k,\Lambda^k), \\
\mathcal{L}_3^k &:= \widetilde{L}_\rho(W^{k+1},H^{k+1},T^k,\Lambda^k), \\
\mathcal{L}_4^k &:= \widetilde{L}_\rho(W^{k+1},H^{k+1},T^{k+1},\Lambda^k),
\end{aligned}
\end{equation*}
so that $\mathcal{L}_1^k = \widetilde{L}^{k}$, while
$\widetilde{L}^{k+1}$ differs from $\mathcal{L}_4^k$ only in the
multiplier.

\emph{$W$-step.} As a function of $W$ alone,
$\widetilde{L}_\rho(\cdot, H^k, T^k, \Lambda^k)$ is a convex quadratic
plus $\tfrac{\mu}{2}\|W\|_F^2$; its Hessian with respect to
$\mathrm{vec}(W)$ is
\begin{equation*}
\rho\,\big(H^k (H^k)^\top \otimes I_m\big) + \mu I
  \ \succeq\ \mu I .
\end{equation*}
It is therefore $\mu$-strongly convex, and since $W^{k+1}$ is its exact
minimizer,
\begin{equation}
\label{eq:Wdec}
\mathcal{L}_1^k - \mathcal{L}_2^k
\ \ge\ \tfrac{\mu}{2}\|\Delta W^k\|_F^2 .
\end{equation}

\emph{$H$-step.} By symmetry, we have 
\begin{equation*}
\mathcal{L}_2^k - \mathcal{L}_3^k
\ \ge\ \tfrac{\mu}{2}\|\Delta H^k\|_F^2 .
\end{equation*}

\emph{$T$-step.} By \ref{ass:smooth}, $\ell$ is $L$-weakly convex,
that is, $\ell(\cdot) + \tfrac{L}{2}\|\cdot\|_F^2$ is convex. Hence the
$T$-subproblem objective, which adds
$\tfrac{\rho}{2}\|T - A^{k+1}\|_F^2$, is $(\rho-L)$-strongly convex.
Since $T^{k+1}$ is its exact minimizer,
\begin{equation*}
\mathcal{L}_3^k - \mathcal{L}_4^k
\ \ge\ \tfrac{\rho-L}{2}\|\Delta T^k\|_F^2 .
\end{equation*}

\emph{$\Lambda$-step.} The Lagrangian is affine in $\Lambda$, so
\begin{equation}
\label{eq:Ldec}
\widetilde{L}^{k+1} - \mathcal{L}_4^k
= \big\langle \Delta\Lambda^k, R^{k+1}\big\rangle
= \tfrac{1}{\rho}\|\Delta\Lambda^k\|_F^2 .
\end{equation}

Summing \eqref{eq:Wdec}--\eqref{eq:Ldec}, where the first three
telescope from $\mathcal{L}_1^k = \widetilde{L}^{k}$ to
$\mathcal{L}_4^k$, and bounding the last using Lemma~\ref{lem:dual}, we obtain 
\begin{equation*}
\widetilde{L}^{k} - \widetilde{L}^{k+1} \ \ge\
\tfrac{\mu}{2}\|\Delta W^k\|_F^2
+ \tfrac{\mu}{2}\|\Delta H^k\|_F^2 
+ \Big(\tfrac{\rho-L}{2} - \tfrac{L^2}{\rho}\Big)
  \|\Delta T^k\|_F^2 .
\end{equation*}
Finally, $\delta > 0$ is equivalent to $\rho^2 - L\rho - 2L^2 > 0$,
whose positive root is $\rho = 2L$; hence $\delta>0$ under
\ref{ass:rho}.
\end{proof}

\paragraph*{Lower boundedness and boundedness of the iterates}

Let us now lower bound the Lagrangian, as well as the sequence of iterates. 

\begin{lemma}[Lower bound on $\tilde L$]
\label{lem:lower}
Under Assumption~\ref{ass:main}, for every $k \ge 0$,
\begin{equation}
\label{eq:lower}
\widetilde{L}^{k+1} \ \ge\ \ell\big(A^{k+1}\big)
+ \tfrac{\rho-L}{2}\,\|R^{k+1}\|_F^2 
+ \tfrac{\mu}{2}\big(\|W^{k+1}\|_F^2 + \|H^{k+1}\|_F^2\big)
\ \ge\ 0 .
\end{equation}
\end{lemma}

\begin{proof}
By Lemma~\ref{lem:dual}, $\Lambda^{k+1} = -\nabla\ell(T^{k+1})$, hence
\begin{equation}
\label{eq:Lexpand}
\widetilde{L}^{k+1} = {}
\ell(T^{k+1})
- \big\langle \nabla\ell(T^{k+1}),\, R^{k+1}\big\rangle
+ \tfrac{\rho}{2}\|R^{k+1}\|_F^2 
+ \tfrac{\mu}{2}\big(\|W^{k+1}\|_F^2+\|H^{k+1}\|_F^2\big).
\end{equation}
The descent lemma applied to $\ell$ between $T^{k+1}$ and $A^{k+1}$
gives
\begin{equation*}
\ell(A^{k+1}) \le {}
\ell(T^{k+1})
- \big\langle \nabla\ell(T^{k+1}),\, R^{k+1}\big\rangle 
+ \tfrac{L}{2}\|R^{k+1}\|_F^2 ,
\end{equation*}
where we used $A^{k+1}-T^{k+1} = -R^{k+1}$. Rearranging the terms gives 
\begin{equation}
\label{eq:descentrearr}
\ell(T^{k+1})
- \big\langle \nabla\ell(T^{k+1}),\, R^{k+1}\big\rangle 
\ge\ \ell(A^{k+1}) - \tfrac{L}{2}\|R^{k+1}\|_F^2 . 
\end{equation}
Substituting \eqref{eq:descentrearr} into \eqref{eq:Lexpand} yields
\eqref{eq:lower}; nonnegativity then follows from \ref{ass:nonneg} and
$\rho > L$.
\end{proof}

\begin{corollary}[Boundedness of the iterates]  
\label{cor:bounded}
Under Assumption~\ref{ass:main}, the sequence $\{\mathcal{X}^k\}$ is
bounded.
\end{corollary}

\begin{proof}
By Lemma~\ref{lem:decrease} the sequence $\{\widetilde{L}^k\}_{k\ge1}$ is
nonincreasing, so $\widetilde{L}^{k} \le \widetilde{L}^{1}$ for all
$k\ge1$. Combining with Lemma~\ref{lem:lower},
\begin{equation*}
\tfrac{\mu}{2}\big(\|W^{k}\|_F^2 + \|H^{k}\|_F^2\big)
  \le \widetilde{L}^{1},
\quad
\tfrac{\rho-L}{2}\|R^{k}\|_F^2 \le \widetilde{L}^{1},
\end{equation*}
so $\{W^k\}$ and $\{H^k\}$ are bounded, hence so is $\{A^k\}$, and
$\{R^k\}$ is bounded. Consequently $T^k = A^k + R^k$ is bounded, and
$\Lambda^{k} = -\nabla\ell(T^{k})$ is bounded by continuity of
$\nabla\ell$. Since the initial iterate is finite, the entire sequence is bounded.
\end{proof}

We emphasize that boundedness of the iterates is \emph{derived} rather
than assumed; the fixed Tikhonov term
$\tfrac{\mu}{2}(\|W\|_F^2+\|H\|_F^2)$ is what makes this possible.

\paragraph*{Subsequential convergence of ADMM to stationary points} We can now prove our main theorem under Assumption~\ref{ass:main}. 

\begin{theorem}
\label{thm:convergence}
Let Assumption~\ref{ass:main} hold and let $\{\mathcal{X}^k\}$ be
generated by \eqref{eq:mod_updates}. Then:
\begin{enumerate}[label=(\roman*), leftmargin=*]
\item $\{\widetilde{L}^k\}_{k\ge1}$ is nonincreasing and converges to a finite
limit;
\item $\|\Delta W^k\|_F$, $\|\Delta H^k\|_F$, $\|\Delta T^k\|_F$ and
$\|\Delta\Lambda^k\|_F$ all tend to $0$;
\item the primal residual vanishes, $\|R^k\|_F \to 0$, so any limit
point is feasible;
\item the sequence has at least one limit point, and every limit point
$\mathcal{X}^\star = (W^\star,H^\star,T^\star,\Lambda^\star)$ satisfies
the first-order stationarity conditions
\begin{equation}
\label{eq:kkt}
T^\star = W^\star H^\star, \quad
\nabla\ell(T^\star) + \Lambda^\star = 0, 
\Lambda^\star (H^\star)^\top = \mu W^\star, \quad
(W^\star)^\top \Lambda^\star = \mu H^\star .
\end{equation}
\end{enumerate}
\end{theorem}

\begin{proof}
(i) By Lemma~\ref{lem:decrease}, $\{\widetilde{L}^k\}_{k\ge1}$ is
nonincreasing; by Lemma~\ref{lem:lower} it is bounded below by $0$.
Hence it converges.

(ii) Summing \eqref{eq:decrease} over $k = 1,\dots,N$ and letting
$N\to\infty$, we get 
\begin{equation*}
\sum_{k=1}^{\infty}\Big(
\tfrac{\mu}{2}\|\Delta W^k\|_F^2
+ \tfrac{\mu}{2}\|\Delta H^k\|_F^2
+ \delta\|\Delta T^k\|_F^2\Big) 
\le\ \widetilde{L}^{1} - \lim_k \widetilde{L}^{k} \ <\ \infty ,
\end{equation*}
so each summand tends to zero. Since $\mu>0$ and $\delta>0$, the first
three claims follow, and
$\|\Delta\Lambda^k\|_F \le L\|\Delta T^k\|_F \to 0$ by
Lemma~\ref{lem:dual}.

(iii) From the dual update, $R^{k+1} = \Delta\Lambda^k/\rho$, and the
right-hand side tends to $0$ by (ii).

(iv) Existence of a limit point follows from
Corollary~\ref{cor:bounded} and the Bolzano--Weierstrass theorem (that is, every bounded sequence has a convergent subsequence). 
Let
$\mathcal{X}^\star$ be the limit of a subsequence indexed by
$\{k_j\}$. By~(ii), the neighbouring iterates share the same limit, so
we may pass to the limit in the optimality conditions of each subproblem.

Feasibility $T^\star = W^\star H^\star$ is (iii). The second condition
in \eqref{eq:kkt} is Lemma~\ref{lem:dual} in the limit, using
continuity of $\nabla\ell$. The optimality condition of the
$W$-subproblem reads
\begin{equation*}
-\big[\Lambda^k + \rho\big(T^k - W^{k+1}H^k\big)\big](H^k)^\top
+ \mu W^{k+1} = 0 .
\end{equation*}
Passing to the limit along $\{k_j\}$ and using
$T^\star = W^\star H^\star$, the bracketed term converges to
$\Lambda^\star$, giving $\Lambda^\star (H^\star)^\top = \mu W^\star$.
The condition for $H$ follows identically from
\begin{equation*}
-(W^{k+1})^\top\big[\Lambda^k
  + \rho\big(T^k - W^{k+1}H^{k+1}\big)\big] + \mu H^{k+1} = 0 .
\end{equation*}
Finally, \eqref{eq:kkt} is precisely the KKT system of
\begin{equation*}
\min_{W,H,T}\ \ell(T)
+ \tfrac{\mu}{2}\big(\|W\|_F^2 + \|H\|_F^2\big)
\ \ \text{s.t.}\ \ T = WH ,
\end{equation*}
whose Lagrangian is $\ell(T) + \langle\Lambda, T-WH\rangle
+ \tfrac{\mu}{2}(\|W\|_F^2+\|H\|_F^2)$.
\end{proof}

}